\documentclass[lettersize,journal]{IEEEtran}
\usepackage[export]{adjustbox}
\usepackage{amsmath}
\usepackage{etoolbox} 
\usepackage{algpseudocode}

\usepackage[dvipsnames,table,xcdraw,usenames,dvipsnames]{xcolor}
\usepackage{soul}
\usepackage{color}

\usepackage{booktabs}
\usepackage{multirow}
\usepackage{siunitx}
\usepackage{lipsum,booktabs}
\usepackage{amssymb}
\usepackage{pifont}
\usepackage{amsmath,amsfonts}
\usepackage{algorithm}
\usepackage{array}
\usepackage[caption=false,font=normalsize,labelfont=sf,textfont=sf]{subfig}
\usepackage{textcomp}
\usepackage{stfloats}
\usepackage{url}
\usepackage{verbatim}
\usepackage{graphicx}
\usepackage{cite}
\usepackage{orcidlink}
\begin{document}
\makeatletter
\newcommand*{\algrule}[1][\algorithmicindent]{\makebox[#1][l]{\hspace*{.5em}\vrule height .75\baselineskip depth .25\baselineskip}}%

\newcount\ALG@printindent@tempcnta
\def\ALG@printindent{%
    \ifnum \theALG@nested>0
        \ifx\ALG@text\ALG@x@notext
            \addvspace{-3pt}
        \else
            \unskip
            \ALG@printindent@tempcnta=1
            \loop
                \algrule[\csname ALG@ind@\the\ALG@printindent@tempcnta\endcsname]%
                \advance \ALG@printindent@tempcnta 1
            \ifnum \ALG@printindent@tempcnta<\numexpr\theALG@nested+1\relax
            \repeat
        \fi
    \fi
    }%
\patchcmd{\ALG@doentity}{\noindent\hskip\ALG@tlm}{\ALG@printindent}{}{\errmessage{failed to patch}}
\makeatother

\title{A Lightweight and Secure Deep Learning Model for Privacy-Preserving Federated Learning in Intelligent Enterprises}

\author{Reza Fotohi\orcidlink{0000-0002-1848-0220}, Fereidoon Shams Aliee\orcidlink{0000-0002-9038-1577}, and Bahar Farahani\orcidlink{0000-0002-7016-6853}
\thanks{Reza Fotohi is with the Faculty of Computer Science and Engineering, Shahid Beheshti University, Tehran 1983969411, Iran (e-mail: r\_fotohi@sbu.ac.ir).}
\thanks{Fereidoon Shams Aliee (Corresponding Author) is with the Faculty of Computer Science and Engineering, Shahid Beheshti University, Tehran 1983969411, Iran (e-mail: f\_shams@sbu.ac.ir).}
\thanks{Bahar Farahani is with the Cyberspace Research Institute, Shahid Beheshti University, Tehran 1983969411, Iran (e-mail: b\_farahani@sbu.ac.ir).}
\thanks{© 2024 IEEE.  Personal use of this material is permitted.  Permission from IEEE must be obtained for all other uses, in any current or future media, including reprinting/republishing this material for advertising or promotional purposes, creating new collective works, for resale or redistribution to servers or lists, or reuse of any copyrighted component of this work in other works.}}


\maketitle
\begin{abstract}
The ever-growing Internet of Things (IoT) connections drive a new type of organization, the Intelligent Enterprise. In intelligent enterprises, machine learning-based models are adopted to extract insights from data. Due to these traditional models' efficiency and privacy challenges, a new federated learning (FL) paradigm has emerged. In FL, multiple enterprises can jointly train a model to update a final model. However, firstly, FL-trained models usually perform worse than centralized models, especially when enterprises' training data is non-IID (Independent and Identically Distributed). Second, due to the centrality of FL and the untrustworthiness of local enterprises, traditional FL solutions are vulnerable to poisoning and inference attacks and violate privacy. Thirdly, the continuous transfer of parameters between enterprises and servers increases communication costs. Therefore, to this end, the $\textsc{FedAnil+}$ model is proposed, a novel, lightweight, and secure \underline{Fed}erated Deep Le\underline{A}r\underline{ni}ng Mode\underline{l} that includes three main phases. In the first phase, the goal is to solve the data type distribution skew challenge. Addressing privacy concerns against poisoning and inference attacks is given in the second phase. Finally, to alleviate the communication overhead, a novel compression approach is proposed that significantly reduces the size of the updates. The experiment results validate that $\textsc{FedAnil+}$ is secure against inference and poisoning attacks with better accuracy. In addition, in terms of model accuracy (13\%, 16\%, and 26\%), communication cost (17\%, 21\%, and 25\%), and computation cost (7\%, 9\%, and 11\%) improvements over existing approaches. The $\textsc{FedAnil+}$ code is available on GitHub\footnote{https://github.com/rezafotohi/FedAnilPlus.}
\end{abstract}


\section{Introduction}
\IEEEPARstart{T}{HE}Internet of Things (IoT) consists of multiple interconnected computing devices and mechanical and digital machines exchanging data with other IoT devices and the cloud. The rapid growth of IoT and cloud computing and the growing volume of data in enterprises have created a big data ecosystem. In this ecosystem, vast volumes of data from different sources are seamlessly integrated and shared among stakeholders. Since sharing and outsourcing data to cloud centers risks breaching privacy and security, a promising technology, Federated learning (FL), has emerged \cite{ref600,ref11}.

FL is a cutting-edge, secure, distributed machine learning (ML) technology that collaboratively trains a shared deep learning model using heterogeneous data from various clients. All client data remains private in FL, and only the updated parameters are sent to the central server \cite{ref6}. This approach bypasses centralized data collection, thereby enhancing security and privacy. Consequently, the trained FL models in local clients must meet several criteria: Achieve better accuracy on non-IID datasets, demonstrate robustness and high resistance to inference and poisoning attacks, and maintain lower communication costs. Thus, this paper addresses the following three significant problems:
\begin{itemize}
\item \emph{Non-IID}: Since enterprises collect training data based on their usage patterns and local environments, \emph{data type distribution skew} often occurs among them. This skew can negatively affect a final global model's accuracy and convergence speed \cite{ref14,ref15,ref244,ref144}.

\item \emph{Privacy concern}: The FL-based technique keeps raw data from local enterprises private. It only shares updated gradient information with the server. However, FL does not guarantee adequate privacy and is vulnerable to poisoning and inference attacks. Model and data poisoning attacks pose significant threats to FL because they aim to degrade the global model's accuracy. Therefore, the attacker injects fake samples into the training dataset in data poisoning. Furthermore, model poisoning attacks manipulate updated parameters, hindering optimization and leading to higher test error rates. In inference attacks, adversaries infer the local sensitive data via the global model parameters to leakage privacy\cite{ref14,ref15,ref244, ref144}.
\item \emph{Communication costs}: FL-based techniques can generate many parameters when building and updating a model. Exchange of these parameters to the server can cause high communication overhead \cite{ref14,ref15,ref244}.
\end{itemize}

Therefore, the $\textsc{FedAnil+}$ model has been proposed to overcome the above three challenges. The main contributions of this research are unfolding as follows:

\begin{itemize}
\item \emph{Non-IID}: In the $\textsc{FedAnil+}$ model, the non-IID challenge is addressed by the heterogeneity in data type distribution skew among different enterprises. To this end, a cosine similarity (CS) and affinity propagation (AP)-based clustering approach is proposed. Therefore, using correct clustering based on these two techniques alleviates the heterogeneity between the local models. Then, the aggregation process can be appropriately performed on homogeneous models in each cluster without reducing the final model's accuracy.
\item \emph{Privacy concern}: Three steps in $\textsc{FedAnil+}$ address privacy concerns. First, to defend against data poisoning attacks, enterprises whose \emph{CS} of the updated gradient vector falls outside the specified range between two thresholds are ignored. If this occurs suspiciously \emph{n} times, the enterprise is removed from the global model update aggregation process. Second, collude attacks are prevented based on the consortium blockchain and by randomly selecting enterprises. Finally, the third step utilizes the Cheon-Kim-Kim-Song (CKKS) \emph{Fully Homomorphic Encryption (FHE)} technique and the blockchain consortium to prevent membership inference attack (MIA), reconstruction, and model poisoning attacks. In this step, local enterprises use \emph{CKKS-FHE} to encrypt the \emph{CHs index vector}, while the local models aggregate without decrypting the model parameters on the server side.
\item \emph{Communication costs}: In $\textsc{FedAnil+}$, a two-step compression method is proposed to solve this challenge. In the first step, k-medoids\cite{ref24}-based quantization is used to overcome the communication costs by entropy reduction in the local gradient vector. The main focus of the quantization technique is to remove gradients that have outliers, noise, or are out of range relative to the gradients of each cluster. The second step uses the entropy coding method in $\textsc{FedAnil+}$. In this step, fewer bits are assigned to gradients with more repetition, and vice versa; more bits are assigned to gradients with less repetition.
\end{itemize}

The other sections are structured as follows. Section \ref{PR1} briefly reviews the preliminary used. The related work is discussed in Section \ref{RW1}. In Section, \ref{FM1}, the details of the $\textsc{FedAnil+}$ are discussed. The convergence analysis is brought in Section \ref{CA1}. In Section \ref{EP1}, experiment results are evaluated. Finally, Section \ref{CC1} brings the conclusions and future work.

\section{Preliminaries}\label{PR1}
The basic preliminaries adopted in $\textsc {FedAnil+}$ are reviewed in this section. The key symbols used are listed in Table \ref{tab2}.

\begin{table}
  \caption{Key notations.}
  \setlength{\tabcolsep}{1.5\tabcolsep}
  \centering
  \begin{tabular}{p{1.2cm}|p{5.2cm}}
    \toprule
    \textbf{Notation} & \textbf{Definition} \\
    \midrule
    $F^k(\boldsymbol\omega)$ & The loss function for enterprise \emph{k}\\
    $F^S(\boldsymbol\omega)$ & The loss function for the server\\
    $DS_{k}$ & Dataset for enterprise \emph{k}\\
    $\boldsymbol\omega^k$ & The local enterprise \emph{k} model\\
    $\boldsymbol\omega^{S}$, $\mathbb{GI}$ & Global model, Global iteration\\
    $\boldsymbol\omega(\tau)^k$ & The local model $\tau$ of enterprise \emph{k}\\
    $\boldsymbol\omega(\tau)^{S}$ & The global model $\tau$ of server\\
    $\overline{\boldsymbol\omega}_{{r}}$ & Averaged model in round \emph{r}\\
    $\Delta c$, $\chi^k$ & Chosen enterprises, Malicious enterprise \emph{k}\\
    \emph{r / R} & Current / Total communication round\\
    $\nabla\boldsymbol\omega$, $\boldsymbol\omega$, \emph{n} & Gradient, Weight, Total enterprises\\
    $\theta^{k}$ & Angle between local and global model\\
    $\Psi$, $\aleph$ & \emph{CH} and their index, Delay of local enterprise \\
    $\Upsilon$ & Initial gradients using the \emph{CH} index\\
    $\tau$, $\mathcal{M}$ & Selected model; ($\tau$ $\in$ $\mathcal{M}$), Models vector\\
    $\kappa$ & Correctly predicted samples\\
    $\iota$ & Total samples in the validation dataset\\
    \bottomrule
  \end{tabular}
  \label{tab2}
\end{table}

\subsection{Cosine Similarity}
According to \eqref{eqCS-Angles}, this technique measures the similarity of two specified non-zero vectors by calculating the angle between two vectors. The output values belong to the interval [$-1$,$1$]. These two values obtained by calculating the angle between two vectors mean that if the output value shows $-1$, it means that the two texts are less similar to each other, but if the output value shows $1$, it means which two texts are more similar to each other \cite{ref11}.
\begin{equation}
\begin{split}
Similarity (x,y)=\cos(\theta)=\\
\frac{x \ast y}{\parallel x \parallel \ast \parallel y \parallel}=\frac{\sum_{i=1}^n x_i \ast  y_i}{\sqrt{\sum_{i=1}^n (x_i)^2 \ast \sum_{i=1}^n (y_i)^2}}.
 \label{eqCS-Angles}
 \end{split}
\end{equation}

In \eqref{eqCS-Angles}, $\boldsymbol{x}$ and $\boldsymbol{y}$ represents the vector $x$ and $y$ respectively. Each $x_i$ and $y_i$ represents an element in these vectors. The angle between two vectors $(x,y)$ is denoted by $\theta$.

The $\textsc{FedAnil+}$ model to measure the similarity between gradient vectors adopts the \emph{CS} technique.

\subsection{Consortium Blockchain}
In a consortium blockchain, some aspects of enterprises are exposed, while others are private. Consensus methods are controlled by predefined nodes. A consortium blockchain is managed by multiple enterprises; Therefore, no single force here has a concentrated result \cite{ref12}.

\subsection{Homomorphic Encryption (HE)}
It is a type of encryption where calculations are performed on encrypted data without initial decryption. Also, the results of the calculations will be encrypted. Formally, such a scheme will be homomorphic if it satisfies \eqref{eq2}\cite{ref6}:
\begin{equation}
\text E(m_{1})*E(m_{2}) = E(m_{1}* m_{2}) \hspace{0.7cm}  \forall m_{1},m_{2} \in M.
 \label{eq2}
\end{equation}

In \eqref{eq2}, Messages and a homomorphic operation are denoted by $M$ and $\ast $, respectively. The main homomorphic operation is described by four main algorithms  ${KeyGen, Enc, Dec}$, and ${Eval}$. These algorithms are listed separately below:

\begin{enumerate}
\item {$KeyGen$($1^{\lambda}$)} $\to$ ($p_{k}$,$s_{k}$): It gets $\lambda$ as the input security parameter and generates a public key $p_{k}$ and a private key $s_{k}$.
\item {$Enc$($p_{k}$,$m$)} $\to$ $c$: It gets $p_{k}$ and $m$ as public key and message, respectively, and generates $c$ as cipher text.
\item {$Dec$($s_{k}$,$c$)} $\to$ $m$: It gets $s_{k}$ and $c$ as private key and cipher, respectively, and gives $m$ as a message.
\item {$Eval$($p_{k}$,$F$,$c_{1}$,$c_{2}$,\ldots,$c_{n}$)} $\to$ $c^{\ast}$: The public key $p_{k}$ takes as input an allowed evaluation function $F$ and computes the cipher texts $c_{1}$ through $c_{n}$ and evaluates to $F$($c_{1}$,\ldots,$c_{n}$) if the following holds true: {$Dec$($s_{k}$,{$Eval$($p_{k}$,$F$,$c_{1}$,$c_{2}$,\ldots,$c_{n}$)})} $=$ $F$($m_{1}$,\ldots,$m_{n}$) where ($c_{1}$,\ldots,$c_{n}$) is the encrypted message ($m_{1}$,\ldots,$m_{n}$).
\end{enumerate}
Informally, the security parameter $\lambda$ represents the difficulty of breaking the encryption key. Generally, an $m$ $\in M$ message can be an integer string or another type of encryption \cite{ref6}. In $\textsc{FedAnil+}$, the \emph{CKKS Fully homomorphic encryption (CKKS-FHE)} has been leveraged to encrypt local models. 

\section{Related Work}\label{RW1}
This section will introduce recent approaches based on privacy-preserving, communication-efficient, non-IID data, and data skews on FL.

In \cite{ref401}, an approach called pFedV is proposed to address feature distribution skew. It modifies the last layer for feature extraction before classification layers, creating variable distribution feature maps instead of compressing the input. In \cite{ref402}, bias among local models is corrected by calibrating the logits to solve label distribution skew. Specifically, in FedBalance, the weak learner is trained locally, and its logits reflect the model's learning ability, which is fully influenced by locally unbalanced data. Merging the logits of the two models reduces the misclassification of minority classes and avoids overlearning of majority classes. In \cite{ref403}, the main skewed task is divided into multiple unskewed (balanced) sub-tasks for quantity distribution skew. Then, the representation of the original task is reconstructed using feature extractors for unskewed sub-tasks. In \cite{ref406}, local models' aggregation and learning operations are performed without access to private data to alleviate privacy concerns. The superiority of the proposed framework over the previous related approaches has been proven in various types of non-IID data distributions in the real world, such as time-skew, quantity-skew, scene-skew, and feature-skew. In \cite{ref407}, to deal with non-IID data skewness, local clients are divided into several groups, and instead of individual models, group models are trained for local clients. The clustering criterion is based on $\textsc{Earth mover's distance}$ to group clients with similar data distribution by measuring the similarities of their models so that each group performs its respective local training in each round.

In \cite{ref2}, a compression framework called sparse ternary compression ({\textsc{STC}}) is proposed, which has low communication overhead. In\cite{ref19}, a Clustered FL ({\textsc{CFL}}) is proposed. The clustering is done between the clients in {\textsc{CFL}}. All the clients are grouped in homogenous clusters based on the similarity criterion. A secure aggregation-based method ({\textsc{RFA}}) is proposed in \cite{ref20} to prevent poisoning attacks. In \cite{ref14}, an FL-based averaging method ({\textsc{FedAvg}}) is proposed, which updates and constructs the global model by using a random selection of clients. {\textsc{FedProx}} emerged as an enhanced version of {\textsc{FedAvg}}, designed to handle non-IID data and improve global model efficiency using Euclidean distance \cite{ref10}. Another notable algorithm is {\textsc{FedAdam}}, introduced for adaptive server optimization, which ensures model convergence despite heterogeneous data \cite{ref16}.

In summary, the $\textsc{FedAnil+}$ model differs from other related approaches in the following areas:

\begin{itemize}
\item According to \cite{ref15}, existing approaches to heterogeneity typically address only one or two aspects, such as label skew, feature skew, temporal skew, and quantity skew, without considering data type skew. However, $\textsc{FedAnil+}$ specifically addresses heterogeneity from the perspective of data type.
\item In the compression step, in the existing approaches, both the vector of the initial gradients and the Cluster Heads have been encoded using Huffman Coding. But in $\textsc{FedAnil+}$ it is done separately: In this way, the initial gradient vector is coded with Adaptive Huffman coding (AHC), and the cluster head vector is encrypted with \emph{CKKS-FHE} and finally, recorded in the blockchain. 
\item In existing methods, K-Means is used for Quantization based on clustering, which has a weakness in cluster head selection and is not sensitive to noisy data. But in {$\textsc{FedAnil+}$}, it is based on K-Medoids, which are sensitive to noisy data and remove noisy data before clustering.
\end{itemize}

\section{$\textsc{FedAnil+}$ Model}\label{FM1}
This section introduces the {$\textsc{FedAnil+}$} model with the following four phases.

\subsection{Overview}
This section first describes the use of blockchain and then the operations repeated in a global model round in $\textsc{FedAnil+}$.
\begin{itemize}
 \item Selecting a simple miner on the blockchain to send initial models to local clients.
\item Random selection of local clients by the simple miner.
\item Initialization and registration of \emph{CKKS-FHE}, $p_{k}$ and $s_{k}$ parameters in the blockchain for selected local clients.
\item Download and update the initial models by local clients and then upload them to the blockchain.
\item Decentralized gradient aggregation to address the single-point-of-failure server problem.
\end{itemize}

In $\textsc{FedAnil+}$, a round of global iteration involves five steps. Each step will be explained in detail below.

\begin{enumerate}
\item {\bf{\textsc{Initialization.}}} Three initial models are created by the central enterprise and uploaded to the blockchain. Then, the parameters of \emph{CKKS-FHE}, $p_{k}$, and $s_{k}$ are set so that local enterprises can download these to update the models.
\item {\bf{\textsc{Selection of Simple and Leader miners.}}} In $\textsc{FedAnil+}$, miners play a crucial role in facilitating heavy blockchain operations. There are two types of miners in the proposed model: Simple and leader miners. The simple miner sends the initial models to local enterprises and, on the server side, evaluates their integrity. The goal of the leader miner is to perform the aggregation operation of the verified local models and update the global model. The simple miner with the highest reward is chosen as the leader miner.
\item {\bf{\textsc{Random selection of enterprises.}}} In this step, the random selection of enterprises prevents collusion attacks among them. By randomly selecting enterprises for each round of training the global model, malicious enterprises cannot predict which ones will be chosen. Consequently, enterprises are unlikely to coordinate collusion attacks with potential partner-enterprises, as the selection process disrupts any patterns that could facilitate such collusion.
\item {\bf{\textsc{Local model training.}}} First, each local enterprise downloads the initial global model from the blockchain using a simple miner. They then train these models in parallel, adapting them based on their respective data types. Next, each enterprise encodes the vector $\Upsilon$ using \emph{AHC} according to \eqref{eq_Privacy-preserving2}, then according to \eqref{eq_Privacy-preserving3}, the vector $\Psi$ is encrypted by \emph{CKKS-FHE}. Finally, it uploads both encoded/encrypted gradient vectors to the blockchain.

\setlength{\parindent}{10pt}
In the $\textsc{FedAnil+}$ to formulate the FDL model, the enterprise \emph{k} data sample is represented as $\bigl($$X_{k}$,$Y_{k}$$\bigl)$, where $Y_{k}$ is the labels and $X_{k}$ is its features. The loss function is defined individually for each enterprise in the training steps. As long as the model is not converged, reducing the loss function is performed. Therefore, in the $\textsc{FedAnil+}$, the dataset of local models is shown as \begin{math}DS_1, DS_2, DS_3,\dots, DS_n\end{math}, where the variable \emph{n} represents the total enterprises. The loss function for the dataset $DS_k$ $\bigl(k\in n\bigl)$, in each local enterprise, is defined, via $F^k(\cdot)$ as \eqref{eq15}: 
\begin{equation}
\begin{split}
F^k\bigl(\boldsymbol\omega(\tau)\bigl)\triangleq \frac{1}{\Big|DS_{k}\Big|} \sum _{k\in DS_k}F\Bigl(E\Bigl(\boldsymbol\omega(\tau)^{k}\Bigl), X_k, Y_k\Bigl)
\label{eq15}
\end{split}
\end{equation}
\begin{equation}
\begin{split}
F\bigl(\boldsymbol\omega(\tau)\bigl)\triangleq \frac{\sum _{_{k\in DS_k}}F\Bigl(E\Bigl(\boldsymbol\omega(\tau)^{k}\Bigl), X_k, Y_k\Bigl)}{\Big|\cup DS_{k}\Big|} =\\ 
\frac{\sum _{k=1}^{n}\Big|DS_{k}\Big|F^k\Bigl(E\Bigl(\boldsymbol\omega(\tau)^{k}\Bigl)\Bigl)}{\Big|\cup DS\Big|}.
\label{eq16}
\end{split}
\end{equation}

In \eqref{eq16}, $DS_{k}$  related to the $\textsc{FedAnil+}$ shows the size of the dataset of enterprises' local models. To specify the entire dataset for all enterprises according to the relation $\big|DS\big|\triangleq \sum _{k=1}^{n}\big|DS_{k}\big|$, and $DS_{k_1}\cap DS_{k_2}=\phi$ for $k_1\neq k_2$.  $F(\boldsymbol\omega(\tau))$ is on the datasets of all enterprises \begin{math}DS_1, DS_2, DS_3,\dots, DS_n\end{math}.

To calculate the loss function by the server, all local enterprises send their loss function along with their dataset size to the server. The purpose of training enterprises models is to calculate $F(\boldsymbol\omega(\tau))$ to obtain the minimum function $F(\boldsymbol\omega(\tau))$ which is shown in \eqref{eq17}:
\begin{equation}
\begin{split}
\boldsymbol\omega(\tau) ^ S=min\biggl\{F(\boldsymbol\omega(\tau))\biggl\}.
\label{eq17}
\end{split}
\end{equation}

In the $\textsc{FedAnil+}$ based on \eqref{eq18}, the Stochastic Gradient Descent (SGD) with momentum is used to optimize the weight of enterprise models. Therefore, each local enterprise shows its local parameters as $\boldsymbol\omega(\tau) _{r}^{k}$ where \begin{math}r=0, 1, 2, 3, \dots, R\end{math} and the variable \emph{r} represents the rounds of local models. All enterprise parameters are initialized at \emph{r} = $0$. At $r \geq 1$, the local model's update is performed on the variable $\boldsymbol\omega(\tau) _{r}^{k}$.
\begin{equation}
\begin{split}
E\Bigl(\boldsymbol\omega(\tau)_{r}^{k}\Bigl)=E\Bigl(\boldsymbol\omega(\tau)_{{r-1}}^{k}\Bigl) - \eta  \nabla F^k\biggl(E\Bigl(\boldsymbol\omega(\tau)_{{r-1}}^{k}\Bigl)\biggl).
\label{eq18}
\end{split}
\end{equation}

In \eqref{eq18}, the variable $\eta$ represents the learning rate. $\nabla F^k\bigl(E\bigl(\boldsymbol\omega(\tau)_{{r-1}}^{k}\bigl)\bigl)$ is the encrypted gradients,  $E\bigl(\boldsymbol\omega(\tau)_{{r-1}}^{k}\bigl)$, for the function \emph{F}. The variable $E\bigl(\boldsymbol\omega(\tau)_{{r}}^{S}\bigl)$, the global parameter is computed by the server using the aggregation operation of all locally updated models of local enterprises ( $E\bigl(\boldsymbol\omega(\tau)_{{r}}^{k}\bigl)$). Also, its volume of data is shown as the weight shown in \eqref{eq19}:
\begin{equation}
\begin{split}
\boldsymbol\omega(\tau)_{_{\sim r}}=\frac{\sum _{k=1}^{n}\Big|DS_{k}\Big|E\bigl(\boldsymbol\omega(\tau)_{{r}}^{k}\bigl)}{\Big|DS\Big|}.
\label{eq19}
\end{split}
\end{equation}

\item {\bf{\textsc{Model aggregation and Block generation.}}} The leader miner, acting as the aggregation server, aggregates the locally encrypted gradient vectors that have passed the \emph{CS} and \emph{AP} steps. Subsequently, the averaged global model is recorded in a new block, enabling the next round of training to be executed by $\Delta c$. This process is repeated until the model achieves better model accuracy.
\end{enumerate}

\subsection{Addressing the non-IID}\label{FM_Unbalanced_non-IID}
In this subsection, the challenge of Data Type distribution skew is addressed according to the following step. 

{\bf{\textsc{Data Type distribution skew.}}} In this step, to solve data type distribution skew, a Personalized FL (PFL) based clustering approach based on \emph{AP} and \emph{CS} is proposed. In PFL, updating and building the models differs from that in FL. In PFL, several models are used, unlike FL. Therefore, each local enterprise trains various global models on their dataset and then sends it to the remote server for aggregation. In $\textsc{FedAnil+}$, this step aims to create different homogenous clusters based on the distribution of different data types to solve the data type distribution skew challenge. Hence, employing the \emph{CS} technique as described in \eqref{eq_Add_Unbalanced_Non-IID1}, the distance between $\boldsymbol\omega(\tau)_{{r}}^{k}$ and $\boldsymbol\omega(\tau)_{{r-1}}^{S}$ is computed and stored in a list. Subsequently, utilizing the \emph{AP} outlined in \eqref{eq_Add_Unbalanced_Non-IID2}, the cluster members are identified based on the \emph{CS} list. As a hyperparameter, the \emph{AP} algorithm does not require the pre-defined total clusters.
\begin{equation}
\begin{split}
\theta/CS \Bigl(E\Bigl(\boldsymbol\omega(\tau)_{{r}}^{k}\Bigl), E\Bigl(\boldsymbol\omega(\tau)_{{r-1}}^{S}\Bigl)\Bigl)=
\\
\frac{\Bigl \langle \Delta E\Bigl(\boldsymbol\omega(\tau)_{{r}}^{k}\Bigl), \Delta E\Bigl(\boldsymbol\omega(\tau)_{{r-1}}^{S}\Bigl) \Bigr \rangle}{\Vert \Delta E\Bigl(\boldsymbol\omega(\tau)_{{r}}^{k}\Bigl) \Vert*\Vert \Delta E\Bigl(\boldsymbol\omega(\tau)_{{r-1}}^{S}\Bigl) \Vert}.
\label{eq_Add_Unbalanced_Non-IID1}
\end{split}
\end{equation}

In \eqref{eq_Add_Unbalanced_Non-IID1}:
\begin{itemize}
\item $\theta$/\emph{CS}: Similarity percentage of two models
\item $\Delta E\bigl(\boldsymbol\omega(\tau)_{{r}}^{k}\bigl)$: The enterprise \emph{k} model.
\item $\Delta E\bigl(\boldsymbol\omega(\tau)_{{r-1}}^{S}\bigl)$: Updated parameters of the previous round of the server.

\end{itemize}
\begin{equation}
\begin{split} \emph{ClusterList[1$\dots$$\mathcal{M}$] = $AP\Bigl(\theta(\tau)_{r}^{k=1},\dots,\theta(\tau)_{r}^{k\in \Delta c}\Bigl)$};
\label{eq_Add_Unbalanced_Non-IID2}
\end{split}
\end{equation}

In \eqref{eq_Add_Unbalanced_Non-IID2}:
\begin{itemize}
\item \emph{ClusterList[1$\dots$$\mathcal{M}$]}: Created clusters.
\item $AP$: Determining cluster members.
\end{itemize}

Finally, {\textsc{FedAvg}} is executed for each cluster, and the global model is built. Since the data type distribution in each cluster group is homogenized using \emph{AP} clustering. As a result, the {\textsc{FedAvg}} aggregation algorithm will have the same performance as it does on homogeneous data. The comprehensive handling of non-IID data and the associated heterogeneous procedure is outlined in Algorithm \ref{alg_Unbalanced_and_Non-IID}.
\begin{algorithm}
\scriptsize
\caption{Non-IID Data in the $\textsc{FedAnil+}$}
\begin{algorithmic}[1]
\Procedure{Heterogeneous}{}{\color{blue}\Comment{\emph{Sect.\hspace{0.03cm}IV(B)}}}
\State {\bf{\textsc{Data\hspace{0.04cm}Type\hspace{0.06cm}Skew();}}}
            \State  $\mathcal{M}=[CNN, ResNet50, GloVe$];
            \For {{ $each$ $\boldsymbol\omega{{(\tau)}}^{k}$} { $\in$$\mathcal{M}$}\hspace{0.12cm} in parallel}
            \For {{ $r=1$ $to$ $R$}}
            \For {{$each$ $k$} {$\in$  $\Delta c$}\hspace{0.12cm}in parallel}
            \State  $\theta(\tau)_{r}^{k} = CS\Bigl(E\Bigl(\boldsymbol\omega(\tau)_{{r}}^{k}\Bigl), E\Bigl(\boldsymbol\omega(\tau)_{{r-1}}^{S}\Bigl)\Bigl)$;
            \State  \emph{ClusterList[1$\dots$$\mathcal{M}$]=$AP\Bigl(\theta(\tau){r}^{k=1},\theta(\tau)_{r}^{k=2},\dots,\theta(\tau)_{r}^{k\in \Delta c}\Bigl)$};
            \State  $E\Bigl(\boldsymbol\omega(\tau)_{r}^{S}\Bigl)\gets FedAvg(\emph{ClusterList[1$\dots$$\mathcal{M}$])}$;
        \EndFor
        \EndFor
        \EndFor
\EndProcedure
\end{algorithmic}
\label{alg_Unbalanced_and_Non-IID}
\end{algorithm}

\subsection{Addressing the Privacy-preserving}\label{FM_Privacy-preserving}
This section prevents inference and poisoning attacks in $\textsc {FedAnil+}$.

{\bf{\textsc{Step 1: Data poisoning attack prevention.}}} On the server side, following the computation of the \emph{CS} between $E\bigl(\boldsymbol\omega(\tau)_{{r-1}}^{S}\bigl)$ and $E\bigl(\boldsymbol\omega(\tau)_{{r}}^{k}\bigl)$, the state of the local models undergoes verification based on a specific condition. The data poisoning attack performs the data poisoning operation using the revealed statistical distribution. Therefore, the desired model is ignored if the angle between two vectors is outside the two thresholds. If this value does not fall between $\varphi_1$ and $\varphi_2$ (as indicated by the green range in Fig. \ref{fig_Fending_off_Data_poisoning}) for five consecutive rounds, the enterprise is classified as malicious and discarded before aggregation process.

According to Fig. \ref{fig_Fending_off_Data_poisoning}, the angle range is between $0$ and $180$ degrees. The cosine similarity technique changes the continuous interval from $+1$ to $-1$ in this range. Using two threshold ranges $\varphi_1$ and $\varphi_2$, these three color ranges are separated as follows:
\begin{figure}[!t]
\centering
\includegraphics[width=2.0in]{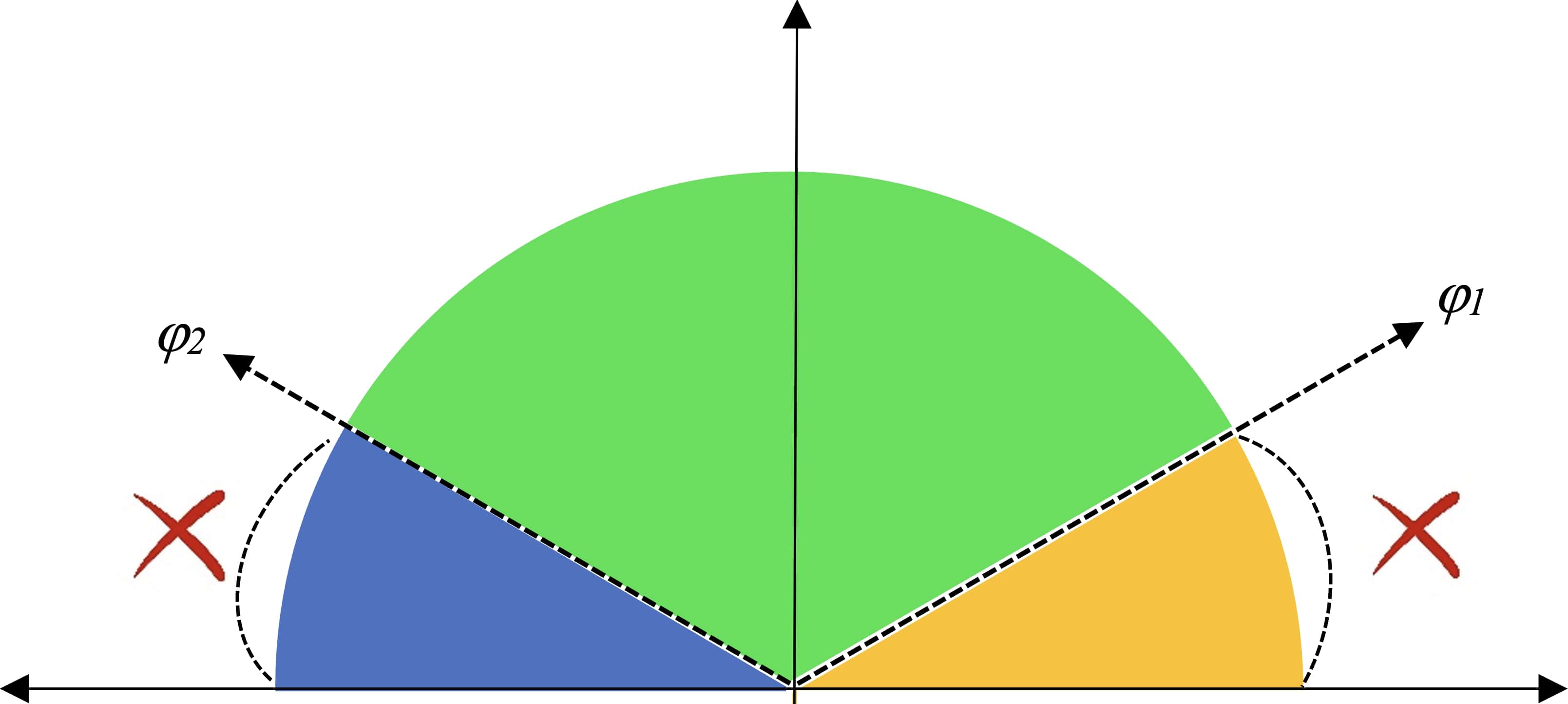}
\caption{$\varphi_1$ and $\varphi_2$ in the $\textsc{FedAnil+}$ model.}
\label{fig_Fending_off_Data_poisoning}
\end{figure}

\begin{itemize}
\item {\bf{\textsc{Yellow.}}} In this range, due to the small cosine angle between the prior round global model and the current local model, the probability of the model being poisoned is high, and therefore, it is discarded.
\item {\bf{\textsc{Blue.}}} By injecting noise into the training data of enterprises, adversaries attempt to induce dissimilar behavior compared to the global model. This results in a large angle (indicating low similarity) between these gradient vectors. Consequently, the local models falling within this range are also excluded from further participation in the model aggregation process to alleviate data poisoning attacks.
\item {\bf{\textsc{Green.}}} Finally, the local models, $E\bigl(\boldsymbol\omega(\tau)_{{r}}^{k}\bigl)$, which form an angle, $\theta(\tau) _{r}^{k}$, within the green range (i.e., between $\varphi_1 = -0.7$ and $\varphi_2 = +0.7$) with the prior global model, $E\bigl(\boldsymbol\omega(\tau)_{{r-1}}^{S}\bigl)$, exhibit a noise-free and normal local model. When a local model is deemed safe by the \emph{CS}, it can be forwarded to subsequent steps.
\end{itemize}

{\bf{\textsc{Step 2: Collude attack prevention.}}} Based on consortium blockchain and random selection of local enterprises, collusion attacks have been prevented. In this model, since $E\bigl(\boldsymbol\omega(\tau)^{S}\bigl)$ represents an average of the behaviors of $E\bigl(\boldsymbol\omega(\tau)_{{r}}^{k}\bigl)\in \Delta c$, selected enterprises must alter their behavior to influence the global model with their attack, termed a collude attack. As the global model serves as an immutable reference and is directly influenced by enterprises, non-local enterprises can potentially bias this model. Therefore, this process is expected to safeguard enterprises from colluding attacks to a significant extent. Moreover, since not all enterprises receive updates of the global model in $\textsc{FedAnil+}$ and the simple miner randomly selects among enterprises, collude attacks can be effectively deterred through random enterprise selection.

{\bf{\textsc{Step 3: Prevention of MIA, Reconstruction, and Model poisoning attacks.}}} To reduce these attacks in $\textsc{FedAnil+}$ model, \emph{CKKS-FHE} and consortium blockchain techniques have been integrated. It should be noted that the server in $\textsc{FedAnil+}$ is honest but curious.

A model poisoning attack seeks to manipulate local parameters. Hence, it's imperative for $\textsc{FedAnil+}$ to prevent the exposure of updated parameters during the aggregation process and global model update. Consequently, in $\textsc{FedAnil+}$, leveraging \emph{CKKS-FHE} can effectively prevent the disclosure of local parameters, enabling servers to execute the requisite computations for aggregation and global model creation. On the other hand, the leader miner aggregates and updates the global model in the consortium blockchain, which has limited access to the public.

In a MIA, the intruder scrutinizes the global model to ascertain whether a specific sample exists within the training dataset. This is done through questions and answers from a trained machine-learning model. At first glance, it seems that to defend against this attack, the model architecture should be hidden from the attackers. However, hiding the model architecture in FL models is impossible because servers and clients follow a model with the same architecture. In fact, in a MIA where the attacker has the architecture used to train the real data, this attacker can, with a fake dataset and, by accessing the parameters of the model update, be able to infer the real training data so that the privacy violation occurs in enterprises. 

In a reconstruction attack, the honest-but-curious server can access local model parameters. Since in the proposed model, all the parameters are encrypted by \emph{CKKS-FHE}, this attack cannot compare the fake data output with any valid source to violate the privacy of the local model's training data by identifying the content of these parameters.

Therefore, to avoid revealing the parameters, the gradients of each enterprise based on two vectors, $\Psi$ and $\Upsilon$ according to \eqref{eq_Privacy-preserving2} and \eqref{eq_Privacy-preserving3} in the blockchain are recorded.
Specifically, first, according to \eqref{eq_Privacy-preserving2}, the vector $\Upsilon$ is encoded using \emph{AHC}. According to \eqref{eq_Privacy-preserving3}, the vector $\Psi$ is encrypted by \emph{CKKS-FHE}, and finally, both vectors are recorded in the blockchain.
\begin{equation}
{{En\Bigl(\rho[\hspace{0.10cm}]\Bigl)}} \longleftarrow Encoding\Bigl(\rho[\hspace{0.10cm}], AHC \Bigl).
\label{eq_Privacy-preserving2}
\end{equation}
\begin{equation}
{{E\Bigl(CH[\hspace{0.10cm}]\Bigl)}} \longleftarrow Encrypt\Bigl(CH[\hspace{0.10cm}], p_{k} \Bigl). \\
\label{eq_Privacy-preserving3}
\end{equation}

In \eqref{eq_Privacy-preserving2} and \eqref{eq_Privacy-preserving3}:
\begin{itemize}
\item $\rho[\hspace{0.10cm}]$, $En\bigl(\rho[\hspace{0.10cm}]\bigl)$: $\Upsilon$, Vector of encoded $\Upsilon$.
\item $AHC$, $p_{k}$: Adaptive Huffman coding, The public key.
\item $CH[\hspace{0.10cm}]$, $E\bigl(CH[\hspace{0.10cm}]\bigl)$: $\Psi$, Vector of encrypted $\Psi$.
\end{itemize}

\subsection{Addressing the Communication costs}\label{FM_Communication costs}
This section describes the gradient compression phase, which includes the following two main steps in detail.

\begin{figure}[!t]
\centering
\includegraphics[width=2.5in, frame]{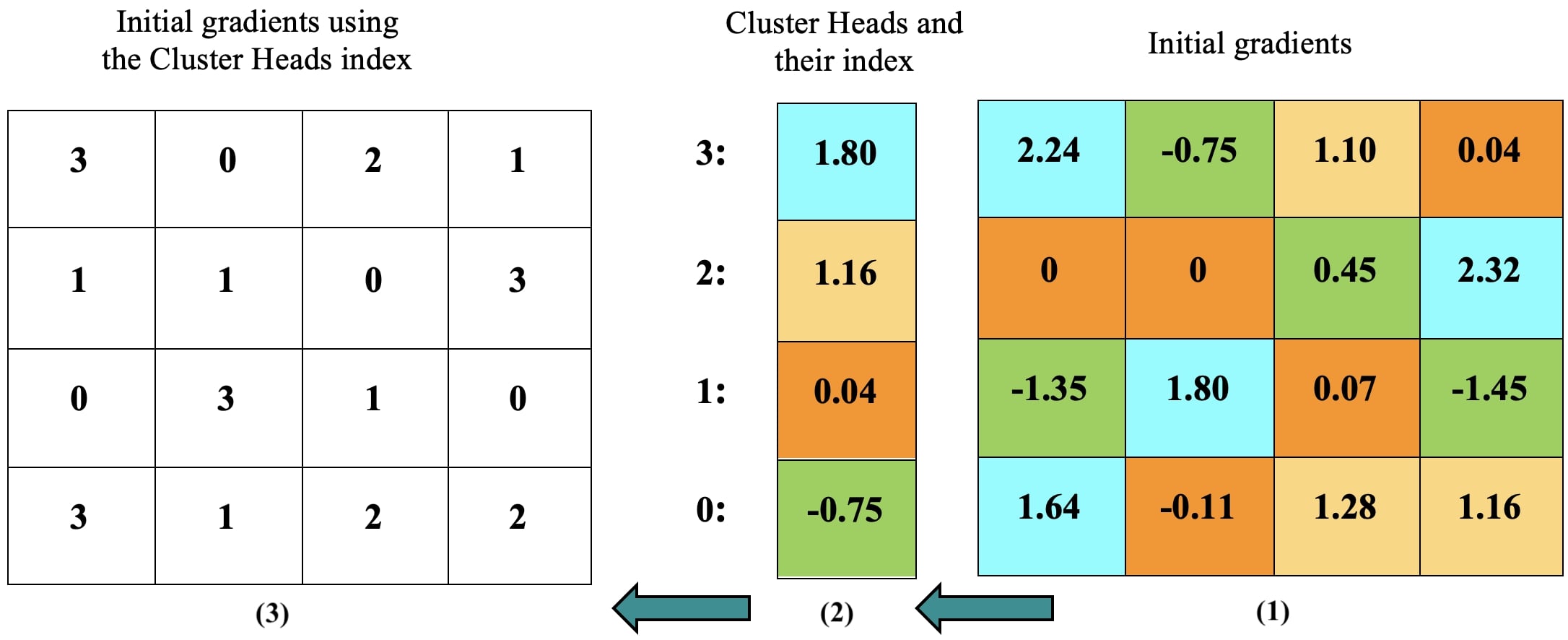}
\caption{Quantization in the $\textsc{FedAnil+}$.}
\label{fig_Quantization}
\end{figure}

{\bf{\textsc{Step 1: Quantization.}}} The quantization technique is used to reduce the entropy in the gradient vector. This technique is based on clustering. Specifically, the \emph{K-Medoids} is used to minimize the number of gradients in the gradient vector, which improves the \emph{K-Means} algorithm. In \emph{K-Medoids}, which operates through iterative repetition, all datasets are segmented into subgroups called clusters. In these clusters, each gradient belongs to only one cluster. For a better understanding, an example of the \emph{Quantization} process is given in Fig. \ref{fig_Quantization}.

{\bf{(1)}}: Fig. \ref{fig_Quantization} shows a $4*4$ matrix containing an enterprise's initial weights. The goal is to minimize the gradients so that similar and close gradients are placed in a cluster.

{\bf{(2)}}: Then, by selecting Medoids on the members of each cluster, a gradient is recorded as \emph{Cluster Head (CH)} in the $\Psi$.

{\bf{(3)}}: Using the $\Psi$ vector, a new matrix called $\Upsilon$ is defined, which includes the index of \emph{CHs} (instead of real gradients).
The vector $\Psi$ is encrypted using \emph{CKKS-FHE}. Encryption protects \emph{CH} gradients against inference and poisoning attacks in the training and aggregation phase. On the other hand, the $\Upsilon$ vector is delivered to the \emph{Lossless Entropy Encoding} step to perform the last step of compression. The reason why \emph{CKKS-FHE} is not applied to this vector is that this vector after compression by \emph{Lossless Entropy Encoding} only contains the number of repetitions of the index related to \emph{CHs}. Therefore, on the server side, no information is revealed by decoding this vector except the number of repetitions of indexes corresponding to \emph{CHs}.

In addition, for calculating $\Psi$ in the \emph{K-Medoids}, the total clusters are automatically calculated using the  \emph{Silhouette index} \cite{ref33}, and the clustering operation is categorized into \emph{K} separate clusters. In the $\textsc{FedAnil+}$, the frequency of optimal clusters was estimated to be $K=5$ based on the  \emph{Silhouette index}.  Also, the \emph{Quantization} operation is shown in \eqref{eq_KMedoids}.
\begin{equation}
\begin{split}
\emph{K-Medoids}\bigl({{\rho [\hspace{0.10cm}]}}\bigl) =
\begin{cases}
 \forall_{{\rho [i]}} \in \Delta c,\hspace{0.2cm}let\hspace{0.2cm}{{\rho [i]}} \longleftarrow i \in CH [\hspace{0.10cm}] \Biggl\}. \\
\end{cases}
\label{eq_KMedoids}
\end{split}
\end{equation}

In \eqref{eq_KMedoids}:
\begin{itemize}
\item ${\rho [i]}$, $CH [\hspace{0.10cm}]$: $\Upsilon$, CHs vector.
\end{itemize}

As a result, by using this \emph{Quantization}, we can greatly reduce the entropy of gradients by clustering and displaying gradients using their \emph{CHs}.

{\bf{\textsc{Step 2: Lossless Entropy Encoding.}}} This step's purpose is to decrease the bits used to transfer the local enterprise's gradient vector. This operation employs \emph{AHC}. The purpose of \emph{AHC} is gradients lossless compression. This algorithm is a type of entropy encoding algorithm. In \emph{AHC}, characters are displayed with a fixed number of bits ($0$, $1$). It allocates fewer bits to frequently used parameters to minimize the required bits and more bits to less frequently used parameters. As described in \textsc{Step 1}, the Quantization output will include two main vectors $\Psi$ and $\Upsilon$. Therefore, in this step, only the $\Upsilon$ vector is encoded using \emph{AHC} and then recorded in the blockchain. On the server side, this vector is decoded by the leader miner, and the gradient vector is reconstructed. \emph{AHC} coding is given in \eqref{eq_Encoding}:
\begin{equation}
\begin{split}
\begin{cases}
 \forall {_k \in \Delta c} ,\hspace{0.2cm}let\hspace{0.2cm}{Encoding \Bigl(\rho[\hspace{0.10cm}], AHC\Bigl)} \Biggl\}. \\
\end{cases}
\label{eq_Encoding}
\end{split}
\end{equation}

The comprehensive handling of communication costs and the associated compression procedure is outlined in Algorithm \ref{alg_Communication_costs}.
\begin{algorithm}
\scriptsize
\caption{Communication costs in the $\textsc{FedAnil+}$}
\begin{algorithmic}[1]
\Procedure{Compression}{}
\State {\bf{Download}} $E\Bigl(\boldsymbol\omega(\tau)_{{r}}^{S}\Bigl)$ {{for $k$} {$\in$  $\Delta c$}};
\State {\bf{\textsc{Quantization();}}} {\color{blue} \Comment{\emph{Sect.\hspace{0.03cm}IV(D),\hspace{0.03cm}Step\hspace{0.03cm}1}}}
\State $CH [\hspace{0.10cm}]$ $\longleftarrow$ $\emph{K-Medoids}\bigl({{\rho [\hspace{0.10cm}]}}\bigl)$;
\State  $E\Bigl(\boldsymbol\omega(\tau)_{{r}}^{k}\Bigl)$ $\longleftarrow$ $\emph{K-Medoids}\bigl({{\rho [\hspace{0.10cm}]}}\bigl)$;
        \For {$each$ $CH [i]$ $\in$  $E\Bigl(\boldsymbol\omega(\tau)_{{r}}^{k}\Bigl)$}\hspace{0.12cm}in parallel
        \For{$each$\hspace{0.09cm}$Cluster\_Mem[j]\in CH[i]$\hspace{0.07cm}in parallel} 
            \State ${{\rho [j]}} \longleftarrow \Bigl\{ i \in CH [\hspace{0.10cm}] \Bigl\}$;
        \EndFor
            \State $E\Bigl(\boldsymbol\omega(\tau)_{{r}}^{k}\Bigl)$ $\longleftarrow$ $\rho [j]$;
        \EndFor
        \State {\bf{\textsc{Lossless\hspace{0.06cm}Encoding();}}} {\color{blue} \Comment{\emph{Sect.\hspace{0.03cm}IV(D),\hspace{0.03cm}Step\hspace{0.03cm}2}}}
        \For {{$each$ $k$} {$\in$  $\Delta c$}\hspace{0.12cm}in parallel}     
        \State {
        $Encoding$ $\Bigl(\rho[\hspace{0.10cm}], AHC \Bigl)$;
        }
        \State {
         $Encryption$ $\Bigl(CH[\hspace{0.10cm}], CKKS-FHE \Bigl)$;
         }
       \EndFor
\State \bf{Upload}$_{k\rightarrow S}$ ${\Bigl(\rho[\hspace{0.10cm}], CH[\hspace{0.10cm}]\Bigl)}$;
\EndProcedure
\end{algorithmic}
\label{alg_Communication_costs}
\end{algorithm}

\section{Convergence analysis}\label{CA1}
To analyze the $\textsc{FedAnil+}$ model convergence, it is assumed that $F(\boldsymbol\omega)$ is non-convex and have two assumptions, which are detailed in \cite{ref144}:

{\bf{\textsc{Assumption 1}}}{$\textsc{($\beta$-smoothness)}$}. Assuming $\nabla F^{k}(\boldsymbol\omega)$ is $\beta$ smoothness, therefore, $\Vert (\nabla F^{k}(\boldsymbol\omega) - \nabla F(\boldsymbol\omega _{\ast}))\Vert \leq \beta \Vert (\boldsymbol\omega^{k} - \boldsymbol\omega _{\ast}) \Vert$, where $\forall$$\boldsymbol\omega^{k}$, $\boldsymbol\omega _{\ast}$ $\in$ $\mathbb{R}^d$. Where $\beta$ is a positive constant.

{\bf{\textsc{Assumption 2.}}} Assuming $F^{k}(\boldsymbol\omega)$ in selected local enterprises be locally convex. Hence, the following formula will hold for $F^{k}(\boldsymbol\omega)$. $F^{k}[(\wp\boldsymbol\omega+(1-\wp)\boldsymbol\omega _{\ast})] \leq [\wp F^{k}(\boldsymbol\omega)+(1-\wp)F(\boldsymbol\omega _{\ast})]$, $\forall$$\boldsymbol\omega$, $\boldsymbol\omega _{\ast}$ $\in$ $\mathbb{R}^d$, $\wp \in [0,1]$ and $\wp$ is a positive constant, and distance for both $\boldsymbol\omega$ and $\boldsymbol\omega _{\ast}$ at a radius $ri > 0$. In the next discussion, we prove the convergence of the $\textsc {FedAnil+}$ model weight parameter $\boldsymbol\omega^{k}$ in training local models.

{\bf{\textsc{Theorem 1.}}} For $\beta$ as a constant, if $\eta\leq \frac{1}{\beta}$, then there exists $\Vert (F^{S}(\boldsymbol\omega_{r+1}) - F^{S}(\boldsymbol\omega _{\ast}))\Vert \leq \Vert (F^{S}(\boldsymbol\omega_{r}) - F^{S}(\boldsymbol\omega _{\ast})) \Vert$, where $F^{S}(\boldsymbol\omega_{r})$ represents the loss function of the aggregated global model. $\boldsymbol\omega_{r}$ and $\boldsymbol\omega _{\ast}$ denote the regular model and the optimized model at round $r$ On the server side, respectively.

{\bf{\textsc{Proof.}}} Following the discussion, the proof of the$\Vert F^{S}(\boldsymbol\omega_{r})-F^{S}(\boldsymbol\omega _{\ast})\Vert^2$ is given.\\\vspace*{0.5em}
$\Vert \bigl(F^{S}(\boldsymbol\omega_{r-1})-\eta\nabla F^{S}(\boldsymbol\omega_{r-1})-F^{S}(\boldsymbol\omega _{\ast})\bigl)\Vert^2$\\
=$\Vert F^{S}(\boldsymbol\omega_{r-1})-F^{S}(\boldsymbol\omega _{\ast})\Vert^2-2\eta\nabla (F^{S}(\boldsymbol\omega_{r-1})^{\mathbb{GI}}F^{S}(\boldsymbol\omega_{r-1})\bigl)\\ \vspace*{0.5em}
-\bigl(F^{S}(\boldsymbol\omega _{\ast}))+\eta^2\Vert\nabla F^{S}(\boldsymbol\omega_{r-1})\Vert^2$ \\ \vspace*{0.5em}
$\leq \Vert F^{S}(\boldsymbol\omega_{r-1})-F^{S}(\boldsymbol\omega _{\ast})\Vert^2-\eta\frac{\Vert\nabla\boldsymbol\omega_{r-1}\Vert^2}{\beta}+\eta^2\Vert\nabla F^{S}(\boldsymbol\omega_{r-1})\Vert^2$\\
=$\Vert F^{S}(\boldsymbol\omega_{r-1})-F^{S}(\boldsymbol\omega _{\ast})\Vert^2-\eta(\frac{1}{\beta}-\eta)\Vert\nabla F^{S}(\boldsymbol\omega_{r-1})\Vert^2$.

In the end, the following relation is output:\\
$\Vert F^{S}(\boldsymbol\omega_{r}) - F^{S}(\boldsymbol\omega _{\ast})\Vert^2 \leq \Vert F^{S}(\boldsymbol\omega_{r-1}) - F^{S}(\boldsymbol\omega _{\ast})\Vert^2$.

{\bf{\textsc{Corollary 1}}}{$\textsc{(anti-inference and anti-poisoning)}$}. If the number of normal enterprises exceeds the number of intruder enterprises, the proposed anti-poisoning and anti-inference approach converges to a model that mirrors the normal enterprise models.

{\bf{\textsc{Proof.}}} To prove the aforementioned Corollary, let's clarify with specific scenarios. In Scenario $1$, all local models are normal. In Scenario $2$, all local models are adversaries. adversary enterprises launch inference and poisoning attacks to disrupt the model's accuracy and privacy. The $\textsc{FedAnil+}$ model will converge in both scenarios $1$ and $2$. However, the direction of the final model obtained from Scenario $1$ and Scenario $2$ will differ. The global models for Scenarios $1$ and $2$ are denoted as $\boldsymbol\omega^{S(normal)}$ and $\boldsymbol\omega^{S(intruder)}$, respectively. The final model obtained after the last round will be a model between $\boldsymbol\omega^{S(normal)}$ and $\boldsymbol\omega^{S(intruder)}$. If the $\mu$ variable represents the percentage of intruder enterprises, the final aggregated model after the $r^{th}$ round is denoted as \eqref{eqCorollary1}:
\begin{equation}
 \boldsymbol\omega^{S}=\Bigr[(1-\mu) \ast \boldsymbol\omega_{r}^{S(normal)}\Bigr]+\Bigr[\mu \ast \boldsymbol\omega_{r}^{S(intruder)}\Bigr].
\label{eqCorollary1}
\end{equation}

According to \eqref{eqCorollary1}, if the majority of local models are normal (i.e., $\mu \approx 0.2$), then the global model will closely resemble $\boldsymbol\omega^{S(normal)}$. In this case, $\mu$ effectively pulls the aggregated model towards the normal enterprises' model, and the proposed anti-poisoning and anti-inference mechanisms mitigate the impact of intruder enterprises in each round.

Next, we elaborate on the aggregated local model's convergence with the same distribution. In the $\textsc{FedAnil+}$ model based on federated learning, there are $n$ local enterprises. The dataset of enterprises is denoted by $DS_1, DS_2, \ldots, DS_n$, each possessing a different data distribution $p^\varrho(\varrho=1,2, \ldots, n)$. Assuming that stochastic gradients $SG(.)$ are unbiased with a distinct probability distribution in each round, i.e., $\mathbb{E}[SG^\varrho(\boldsymbol\omega_{r})]=\nabla F^\varrho(\boldsymbol\omega_{r})$.

{\bf{\textsc{Theorem 2.}}} In the $\textsc{FedAnil+}$ model, following the addressing of the non-IID challenge (as discussed in Section \ref{FM_Unbalanced_non-IID}), it is assumed that the set of uploaded weight parameters is selected from datasets with the same distribution $p^\varrho$. We can establish the following relation compared to $\textsc{FedAvg}$:
\begin{equation}
 \mathbb{E}\Vert \boldsymbol\omega_{r}^{\varrho} - \boldsymbol\omega _{\ast}^{\varrho} \Vert^2 \leq \mathbb{E}\Vert \Bar{\boldsymbol\omega}_{r} - \boldsymbol\omega _{\ast}^{\varrho} \Vert^2,
\label{eqTheorem2}
\end{equation}

 where $\boldsymbol\omega _{\ast}^{\varrho}$ represents the optimized weight with $p^\varrho$ distribution to fit the dataset. $\boldsymbol\omega_{r}^\varrho$ is the received local model with $p^\varrho$ distribution. And finally, $\bar{\boldsymbol\omega}_{r}$ defines the {$\textsc{FedAvg}$} uniform global model in round $r$.

{\bf{\textsc{Proof.}}} Utilizing induction, we can prove the result. First, we include the following two relationships:
\begin{equation}
  \bar{\boldsymbol\omega}_{_{r=1}}=\boldsymbol\omega_{0}-\eta \nabla \bar{SG}_{r=1},
\label{eqProo2}
\end{equation}
\begin{equation}
 {\boldsymbol\omega}_{_{r=1}}^\varrho=\boldsymbol\omega_{0}-\eta \nabla {SG}_{r=1}^\varrho,
\label{eqProo3}
\end{equation}

 After the first round on the server side, {$\textsc{FedAvg}$} gradients are displayed with $\bar{SG}_{r=1}$ and {$\textsc{FedAnil+}$} gradients with ${SG}_{r=1}^\varrho$. The execution of the first round of {$\textsc{SGD with momentum}$} with {$\textsc{FedAvg}$} is given in \eqref{eqProo2} and the execution of the first round of {$\textsc{SGD with momentum}$} with {$\textsc{FedAnil+}$} is given in \eqref{eqProo3}. Therefore, according to \eqref{eqProo2} and \eqref{eqProo3}, we can conclude \eqref{eqProo4}:
\begin{equation}
 \mathbb{E}\Vert \boldsymbol\omega_{r=1}^\varrho - \boldsymbol\omega _{\ast}^{\varrho} \Vert^2 \leq \mathbb{E}\Vert  \bar{\boldsymbol\omega}_{r=1} - \boldsymbol\omega _{\ast}^{\varrho} \Vert^2.
\label{eqProo4}
\end{equation}

 After this, it is assumed that \eqref{eqProo5} is correct in $r^{th}$ round, we will have:
\begin{equation}
 \mathbb{E}\Vert \boldsymbol\omega_{r}^\varrho - \boldsymbol\omega _{\ast}^{\varrho} \Vert^2 \leq \mathbb{E}\Vert  \bar{\boldsymbol\omega}_{r} - \boldsymbol\omega _{\ast}^{\varrho} \Vert^2.
\label{eqProo5}
\end{equation}

 Now, using the \eqref{eqProo2} and \eqref{eqProo3}, we check the round $(r+1)^{th}$ and then the \eqref{eqProo6} is obtained:
\begin{equation}
 \mathbb{E}\Vert \boldsymbol\omega_{r}^\varrho - \eta \nabla \bar{SG}_{r} - \boldsymbol\omega _{\ast}^{\varrho} \Vert^2 \leq \mathbb{E}\Vert  \bar{\boldsymbol\omega}_{t} - \eta \nabla {SG}_{r}^\varrho - \boldsymbol\omega _{\ast}^{\varrho} \Vert^2,
\label{eqProo6}
\end{equation}

 And finally, the \eqref{eqProo7} can be expressed:
\begin{equation}
 \mathbb{E}\Vert \boldsymbol\omega_{r+1}^\varrho - \boldsymbol\omega _{\ast}^{\varrho} \Vert^2 \leq \mathbb{E}\Vert  \bar{\boldsymbol\omega}_{t+1} - \boldsymbol\omega _{\ast}^{\varrho} \Vert^2.
\label{eqProo7}
\end{equation}

When the optimized parameters of local models are correctly compatible with the dataset pattern of local enterprises, we call it convergence. The convergence has been accepted if the training parameters undergo a constant and unchanged process. The ultimate goal of convergence in non-IID data is to converge each non-IID data set to the optimal model individually. Therefore, if the loss function $F(\boldsymbol\omega)$ approaches $0$ continuously, the trained model converges to the optimal model.

\section{Experiments}\label{EP1} This section compares the performance of the $\textsc{FedAnil+}$ model against six well-known baseline methods: {\textsc{FedAvg}}\cite{ref14}, {\textsc{FedProx}}\cite{ref10}, {\textsc{FedAdam}}\cite{ref16}, {{\textsc{STC}}\cite{ref2}, {\textsc{CFL}}\cite{ref19}, and {\textsc{RFA}}\cite{ref20}.

\subsection{Experimental Setup}
The experiments rely on a statistical concept called the Dirichlet distribution (denoted as Dir ($\alpha$)), which is a probability distribution used for continuous, multi-dimensional data. This distribution is characterized by a parameter $\alpha$, which must be a positive number greater than zero ($p \thicksim Dir (\alpha), \alpha > 0$) \cite{ref7}. Dir ($\alpha$) is used to generate non-IID datasets. The parameter $\alpha$ is utilized to control the level of non-IID data distribution for each enterprise. As illustrated in Fig. \ref{fig_Dricklet}a, a lower $\alpha$ value leads to a more skewed distribution, resulting in a higher degree of non-IID data. Conversely, a larger $\alpha$ value creates a distribution closer to a uniform one, mimicking IID (Fig. \ref{fig_Dricklet}c). The $\textsc{FedAnil+}$ model utilizes the Dirichlet distribution (with $\alpha$ set to $0.1$) to partition the overall dataset. This results in each participating enterprise having a unique distribution of class types within their local dataset. The total samples in each local dataset will also vary. Therefore, the $\textsc{FedAnil+}$ model considers one mode for non-IID data called \emph{Data Type skew}, in which the data type differs in different enterprises.
\begin{figure}[!t]
\centering
\includegraphics[width=3.5in]{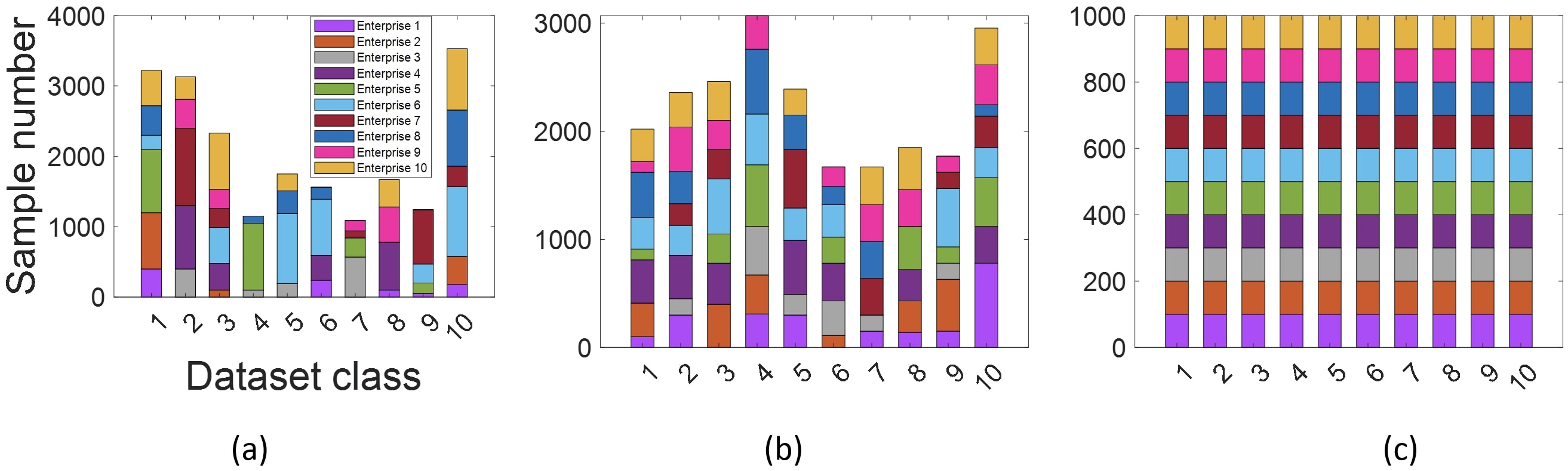}
\caption{Non-IID data distribution. Different colors represent different labels, and each column represents an enterprise's data distribution. (a) $\alpha \rightarrow 0.1$;  (b) $\alpha \rightarrow 1$; (c) $\alpha \rightarrow\infty$.}
\label{fig_Dricklet}
\end{figure}

{\bf{\textsc{Implementation.}}} Our experiments were conducted on the macOS Ventura operating system using Python $3.10.9$. We utilized the PyTorch \cite{ref25} library to train the models. Performance evaluation of the $\textsc{FedAnil+}$ model has been done on the four popular datasets. According to \cite{ref14}, used hyperparameters for $\textsc{FedAnil+}$ model are shown in Table \ref{tab3}.
\begin{table}
  \caption{Hyperparameters.}
  \setlength{\tabcolsep}{1.7\tabcolsep}
  \centering
  \begin{tabular}{ p{3.0cm}|p{4.6cm}}
    \toprule
    \textbf{Hyperparameters} & \textbf{Description} \\
    \midrule
    \emph{SGD with momentum=$0.9$} & Optimizer for CNN\\
    \emph{Adam$\bigl(\beta_{_1},\beta_{_2}=0.9\bigl)$} & Optimizer for ResNet50\\
    $K=5$ & Total clusters for \emph{K-Medoids}\\
    $R=50$ & Total communication rounds\\
    $B_1=64$ & Training batch size for local models\\
    $B_2=128$ & Testing batch size for local models\\
    $\mu=20\%$, $\varepsilon=30$ & Rate of Malicious, Total \emph{Epochs}\\
    $C=100$, $\eta=0.01$ & Total enterprises, Learning rate\\
    \bottomrule
  \end{tabular}
  \label{tab3}
\end{table}

{\bf{\textsc{Datasets and Models.}}} To evaluate the $\textsc{FedAnil+}$, we will have four diverse non-IID datasets and three machine-learning models according to Table \ref{tab4}.

\begin{enumerate}
\item \emph{Sentiment analysis.} The purpose of the Sent140 \cite{ref28} is sentiment analysis. A linear model using average $\textsc{GloVe embeddings}$ \cite{ref29} of tweet words was used. Also, the binary logistic loss was used to train the model. Sent140 has $2$ classes. Table \ref{tab4} gives the rest of its more details.
\item \emph{Image classification.} The purpose of the Fashion-MNIST \cite{ref65} is image classification. This dataset contains low-resolution grayscale images designed with a scale of $28*28$. Fashion-MNIST has $10$ classes. Table \ref{tab4} gives the rest of its more details.
\item \emph{Handwritten character recognition.} The purpose of the FEMNIST dataset \cite{ref27} is to analyze handwritten characters. This dataset contains grayscale images designed with a scale of $28*28$. FEMNIST has $62$ classes. Table \ref{tab4} gives the rest of its more details.
\item \emph{Image classification.} The purpose of the CIFAR-10 \cite{ref30} is image classification. This dataset contains color images designed with a scale of $32*32$. CIFAR-10 has $10$ classes. Table \ref{tab4} gives the rest of its more details.

\end{enumerate}
\begin{table}
  \caption{The dataset used in the $\textsc{FedAnil+}$ (Sentiment Analysis=S.A, Image Classification=I.C, Character-level=C.L).}  
  \setlength{\tabcolsep}{1.7\tabcolsep}
  \centering
  \begin{tabular}{p{1.9cm}|p{0.8cm}|p{1.0cm}|p{0.6cm}|p{0.6cm}}
    \toprule
    \textbf{Dataset} & \textbf{Task} & \textbf{Models}& \textbf{\#Train}& \textbf{\#Test}\\
    \midrule
    Sent140 & S.A & GloVe & $57K$ & $15K$\\
    Fashion-MNIST & I.C & CNN & $60K$ & $10K$\\
    FEMNIST & C.L & CNN & $49K$ & $4.9K$\\
    CIFAR-10 & I.C & ResNet50 & $50K$ & $10K$\\
    \bottomrule
  \end{tabular}
  \label{tab4}
\end{table}

{\bf{\textsc{Evaluation Metrics.}}} The metrics used to evaluate the proposed model are explained in detail below:

\begin{enumerate}
\item \emph{Accuracy.} In $\textsc{FedAnil+}$, a model's performance is measured by how well it predicts on a validation dataset. This is calculated as the number of correct predictions ($\kappa$) divided by the total samples ($\iota$) in the validation set, as shown in \eqref{eq6}.
\begin{equation}
 \emph{Accuracy} = \frac {\kappa}{{\iota}}*100.
    \label{eq6}
\end{equation}

\item \emph{Communication overhead.} One of the important goals of $\textsc {FedAnil+}$ is to achieve the highest compression rate to reduce communication costs. To achieve this goal, the communication cost was calculated using gradient vector compression and reducing the bits required to transfer the local gradient vector of each enterprise to the server. This metric is calculated by \eqref{eq5-1}, \eqref{eq5-2}, and \eqref{eq5-3}.
\begin{equation}
 \emph{COMM}_{_{C2S\_Side}}= \sum _{r=1}^{R}\biggl(\sum _{k=1}^{\Delta c} {b}_{_{\rho[\hspace{0.07cm}]}} + {b}_{_{CH[\hspace{0.07cm}]}}\biggl).
 \label{eq5-1}
\end{equation}
 \begin{equation}
 \emph{COMM}_{_{S2C\_Side}}= \sum _{r=1}^{R}\biggl({b}_{{E\bigl(\boldsymbol\omega(\tau)^{S}\bigl)}}\biggl).
\label{eq5-2}
\end{equation}
 \begin{equation}
 \begin{split}
\emph{COMM}_{_{Total}}=\biggl(\emph{COMM}_{_{C2S\_Side}}+\emph{COMM}_{_{S2C\_Side}}\biggl).
\end{split}
\label{eq5-3}
\end{equation}
In \eqref{eq5-1} and \eqref{eq5-2}:
\begin{itemize}
\item \emph{${{COMM}_{_{C2S\_Side}}}$}: The communication costs from local enterprises to the server. 
\item \emph{${{b}_{_{\rho[\hspace{0.07cm}]}}}$}: The number of bits used by each local enterprise in entropy coding (\emph{AHC}) on the $\Upsilon$ vector.
\item \emph{${{b}_{_{CH[\hspace{0.07cm}]}}}$}: The consumed bits of each local enterprise in encryption on the $\Psi$ vector.
\item \emph{${{COMM}_{_{S2C\_Side}}}$}: The communication costs from the server to the local enterprises. 
\item \emph{${b}_{{E\bigl(\boldsymbol\omega(\tau)^{S}\bigl)}}$}: Bits used for global model parameters.
\item \emph{${COMM}_{_{Total}}$}: Total number of bits used.
\end{itemize}

\item \emph{Computation overhead.} The computation overhead of the $\textsc {FedAnil+}$ model was calculated separately for the enterprises and server sides. The techniques that inject computation cost for the $\textsc{FedAnil+}$ on the enterprise side include: \emph{Quantization}, \emph{Coding}, \emph{Training}, and \emph{FHE}. On the server side, include \emph{AP} and \emph{CS}. The key point is that the \emph{FHE} on the client side has the highest computation, with a complexity of \emph{O($N^2$)}. On the server side, the \emph{AP} and \emph{CS} create the most computation. This is because they are done sequentially, one after the other, resulting in a computation overhead of \emph{O($N^2$)}. Finally, the highest computation cost is related to the server, which is of the order of \emph{O($N^2$)}.
\end{enumerate}

\subsection{Experimental Results}
In this section, we train three models on four diverse datasets, Sent140 \cite{ref28}, Fashion-MNIST \cite{ref65}, FEMNIST \cite{ref27}, and CIFAR-10 \cite{ref30}, and evaluated the robustness of the $\textsc{FedAnil+}$ model. Then, the $\textsc{FedAnil+}$ model was compared and evaluated against other approaches, {\textsc{STC}}, {\textsc{CFL}}, {\textsc{RFA}}, {\textsc{FedAdam}}, {\textsc{FedProx}}, and {\textsc{FedAvg}}. These comparisons were done according to Fig. \ref{fig_ALL_accuracy}, Fig. \ref{fig_ALL_communication}, and Fig. \ref{fig_ALL_computation}. 

{\bf{\textsc{Overall Accuracy.}}} As illustrated in Fig. \ref{fig_accuracy_Sent140_case}, Fig. \ref{fig_accuracy_Fashion-MNIST_case}, Fig. \ref{fig_accuracy_FEMNIST_case}, and Fig. \ref{fig_accuracy_CIFAR10_case}, with the increase in the number of enterprises from $40$ to $100$, the $\textsc{FedAnil+}$ accuracy has increased. The first reason for the improved accuracy of the $\textsc{FedAnil+}$ in later training rounds is that on the server side, the models from all enterprises are clustered based on the distribution of their \emph{data type}, using clustering techniques like \emph{CS} and \emph{AP}. Then, the aggregation is performed within these homogeneous clusters. Therefore, the \emph{CS} and \emph{AP} techniques aim to cluster the heterogeneous models from the local models. The homogenization of the local models, achieved through the clustering, reduces convergence time and increases global model accuracy. The second reason for the improved accuracy of the $\textsc{FedAnil+}$ model is that it helps prevent poisoning attacks. These attacks aim to increase the convergence time and reduce the accuracy of the global model. The reason behind the low accuracy of the {\textsc{CFL}} approach is that they use an encryption technique, which leads to a significant loss of gradients during the decryption process (referred to as the Lossy data approach). This gradient loss ultimately results in a lower model accuracy. The efficiency of the {\textsc{RFA}} approach is weak on non-IID data, which leads to low model accuracy. A strong reason for the poor accuracy of the model in the {\textsc{STC}} method is the use of the gradient compression technique, which has led to the loss of useful gradients. In {\textsc{FedAdam}}, {\textsc{FedAvg}}, and {\textsc{FedProx}}, the performance against poisoning attack and non-IID data is weak, causing the model to diverge. Therefore, these have caused the model's accuracy to decrease.
\begin{figure*}[!t]
\centering
\subfloat[]{\includegraphics[width=1.2in]{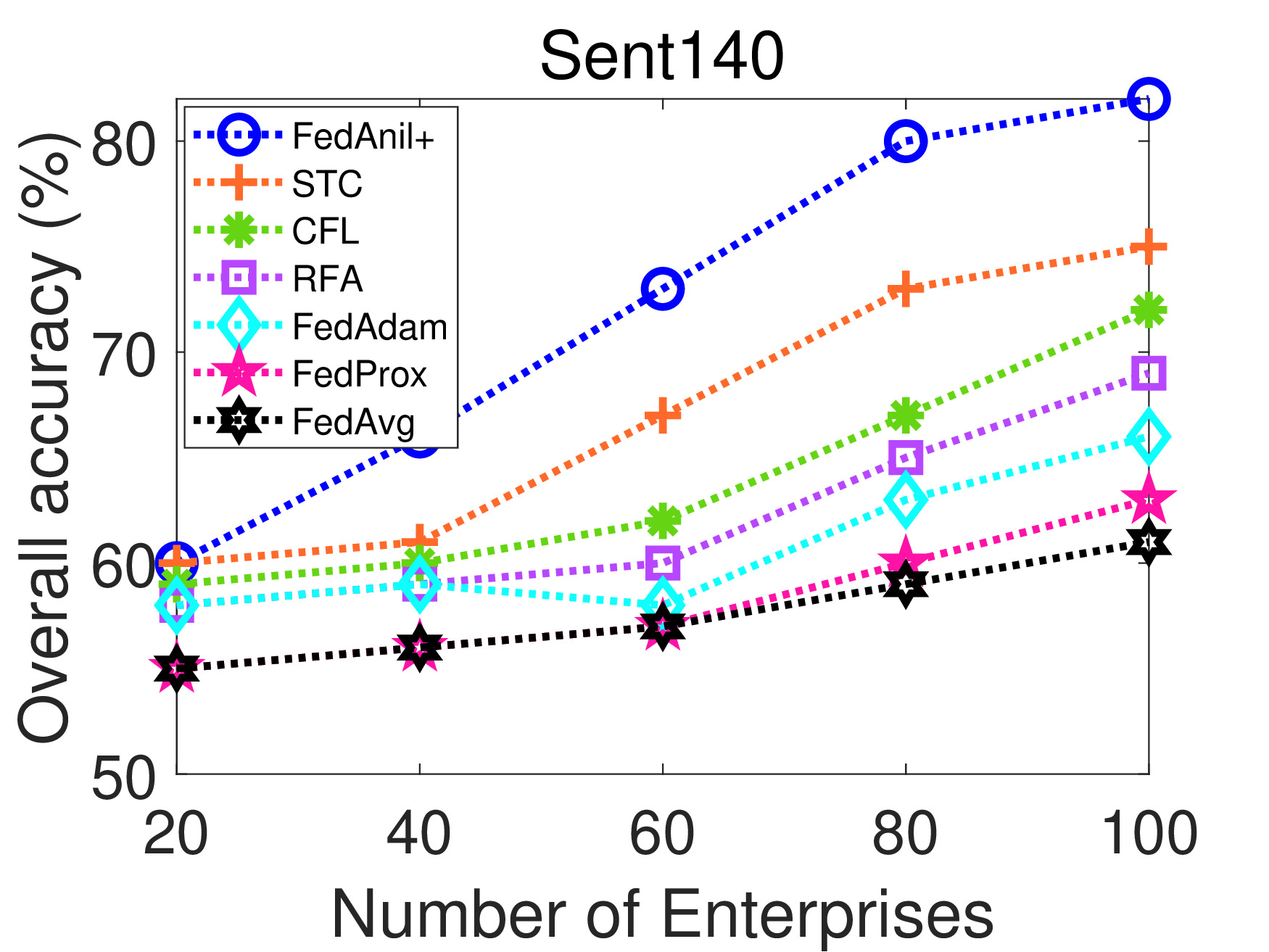}%
\label{fig_accuracy_Sent140_case}}
\hfil
\subfloat[]{\includegraphics[width=1.2in]{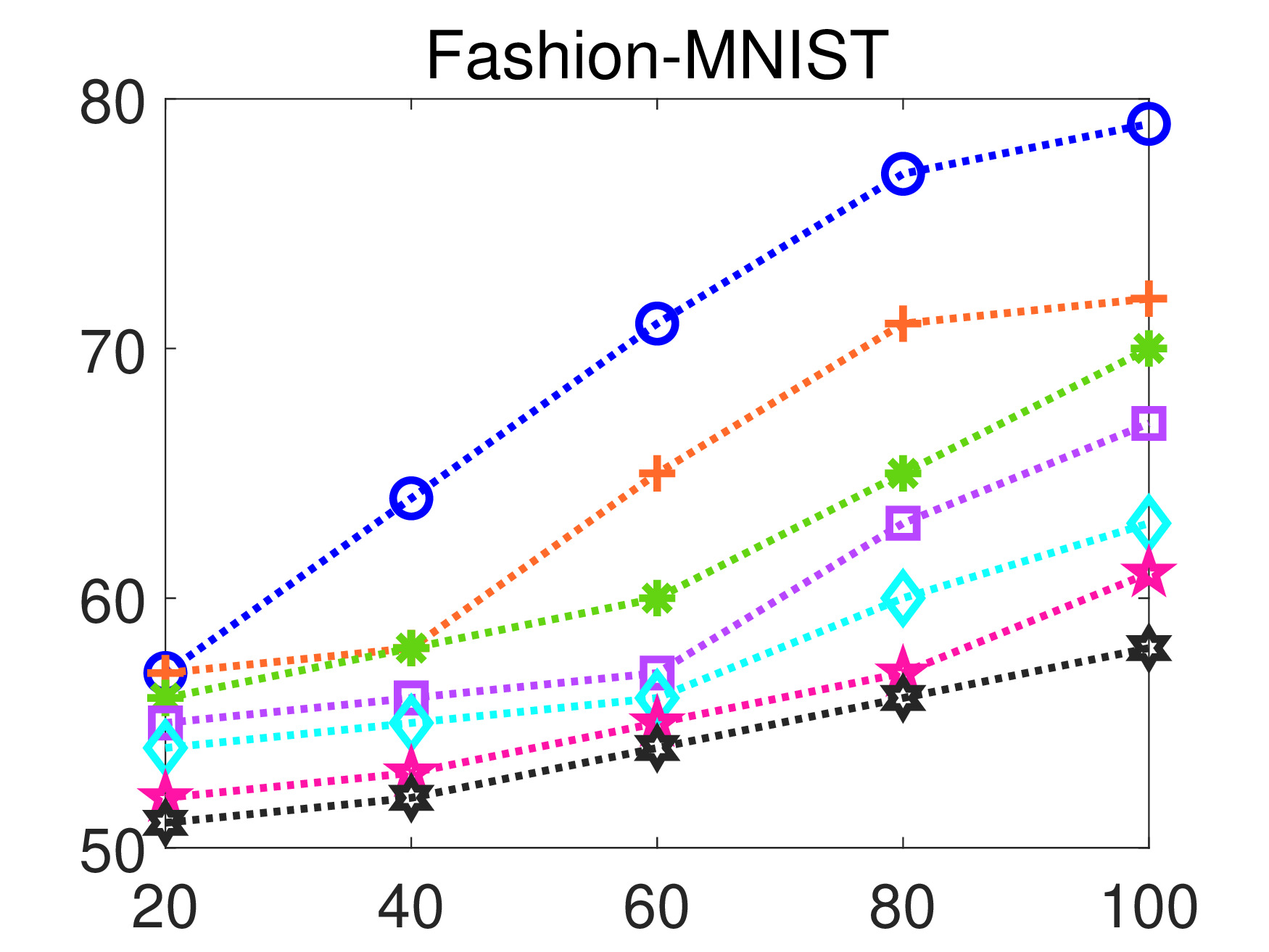}%
\label{fig_accuracy_Fashion-MNIST_case}}
\hfil
\subfloat[]{\includegraphics[width=1.2in]{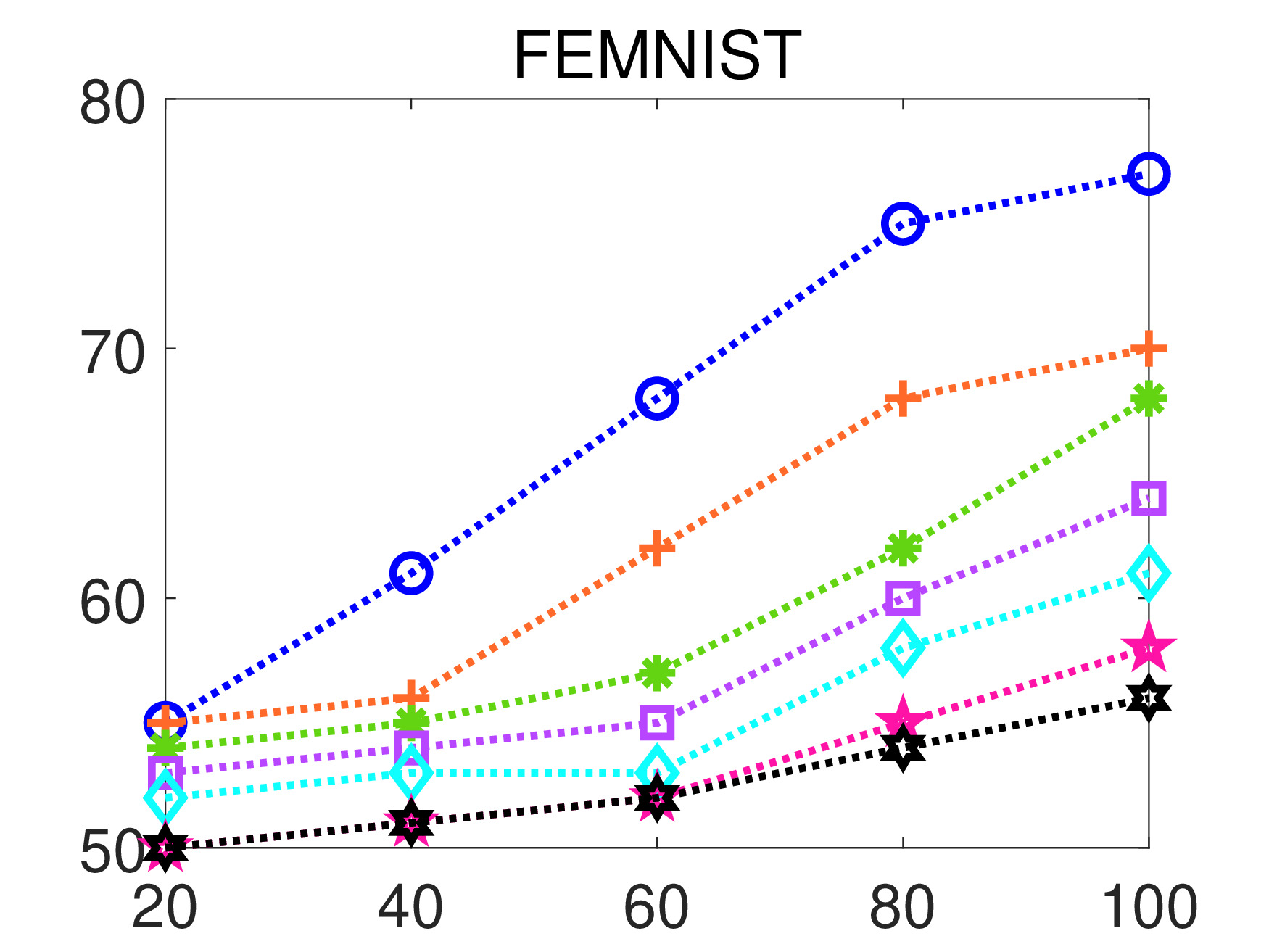}%
\label{fig_accuracy_FEMNIST_case}}
\hfil
\subfloat[]{\includegraphics[width=1.2in]{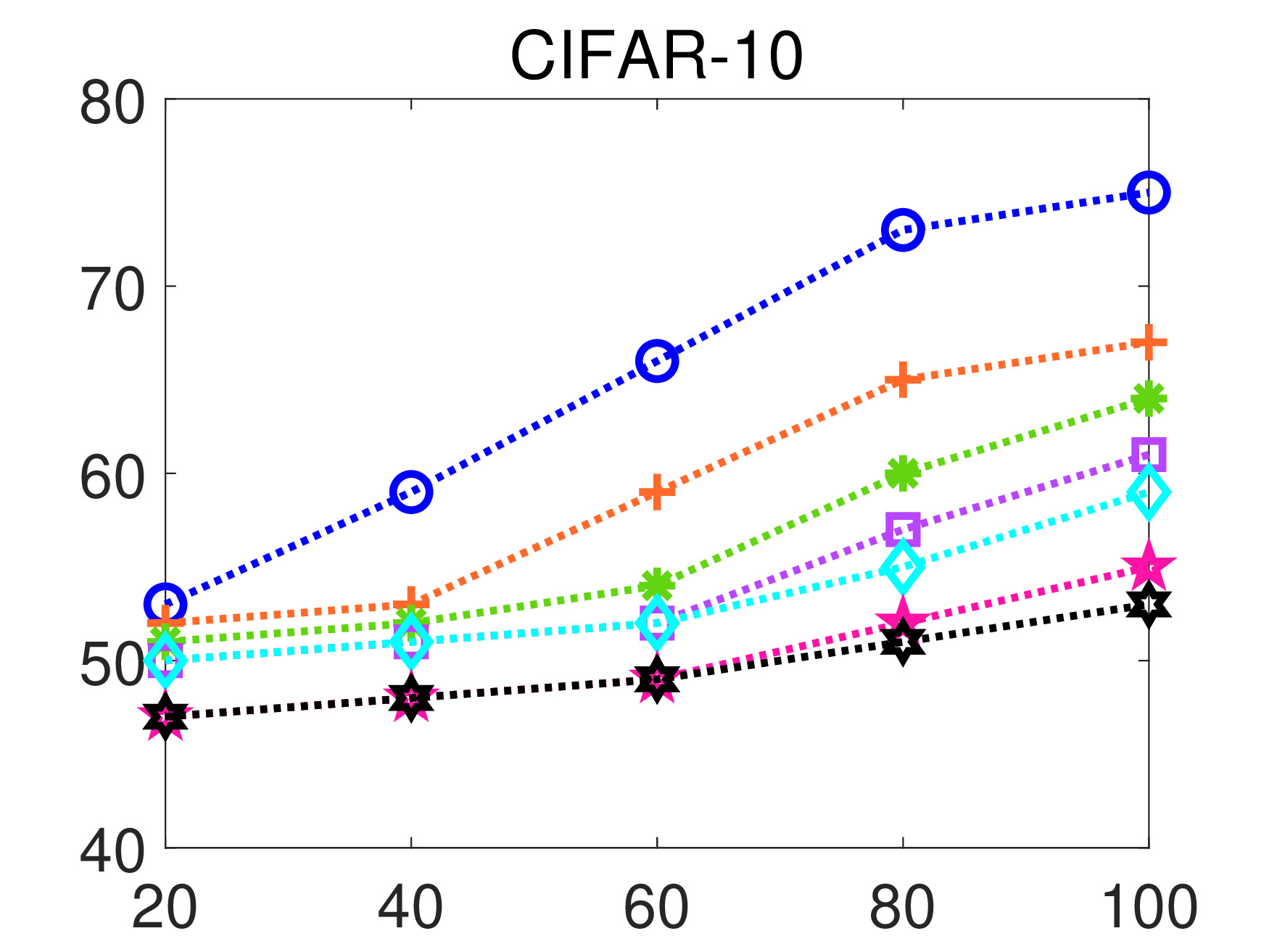}%
\label{fig_accuracy_CIFAR10_case}}
\hfil
\caption{Comparison of the overall accuracy between $\textsc{FedAnil+}$ and existing methods where the $\varepsilon$ is fixed at $30$ and $R=50$. ($\mu=20\%$).}
\label{fig_ALL_accuracy}
\end{figure*}

{\bf{\textsc{Communication Overhead.}}} As shown in Fig. \ref{fig_communication_Sen140_case}, Fig. \ref{fig_communication_Fashion-MNIST_case}, Fig. \ref{fig_communication_FEMNIST_case}, and Fig. \ref{fig_communication_CIFAr10_case}, the total communication cost of the $\textsc{FedAnil+}$ model has been compared and evaluated. While the communication overhead increased for all approaches, the communication cost for the FedAnil+ model was much smaller than the other approaches. In the $\textsc{FedAnil+}$ model, due to the use of the \emph{K-Medoids Quantization} and \emph{LossLess Entropy Encoding}, more bits are compressed and just useful gradients are transmitted to the aggregator server. In baseline approaches, by injecting poisoning attacks, the model diverges, and its accuracy decreases. Because the model accuracy is affected by poisoning attacks, local enterprises and servers consume more rounds. As a result, more bits are consumed, and this work has caused a lot of communication overhead for the models. Because the parameter compression operation in the {\textsc{STC}} approach is performed after the model training steps, the quantization process and the communication cost are not optimized. In the {\textsc{CFL}} approach, Bipartition is used to find a correct partitioning, which, in addition to having heavy operations, this approach is not based on parameter update compression and has a higher communication overhead. The {\textsc{RFA}} approach performs poorly in compression operations on non-IID data. Also, due to not using entropy coding, the length of bits is longer for transferring gradients.
\begin{figure*}[!t]
\centering
\subfloat[]{\includegraphics[width=1.2in]{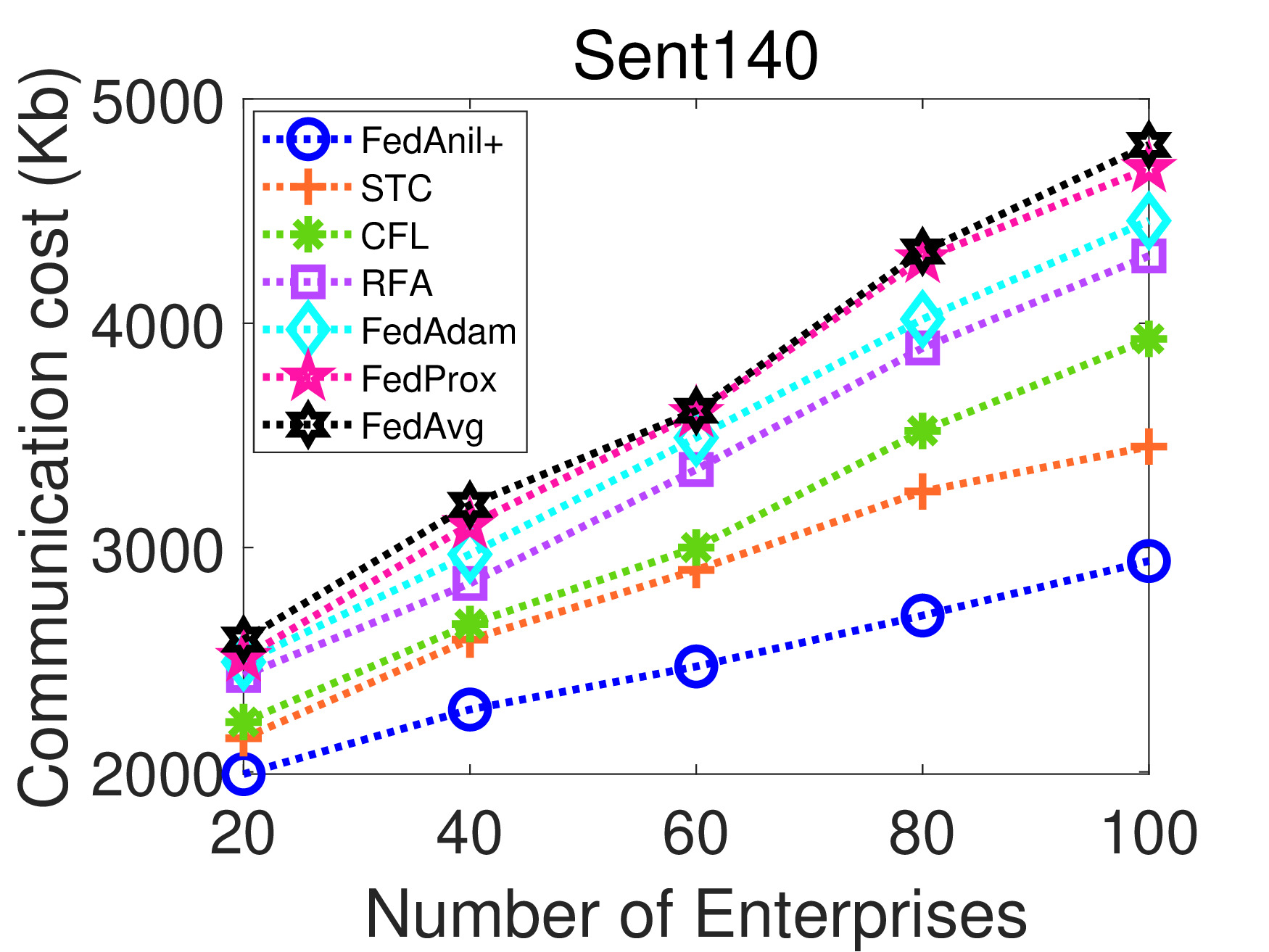}%
\label{fig_communication_Sen140_case}}
\hfil
\subfloat[]{\includegraphics[width=1.2in]{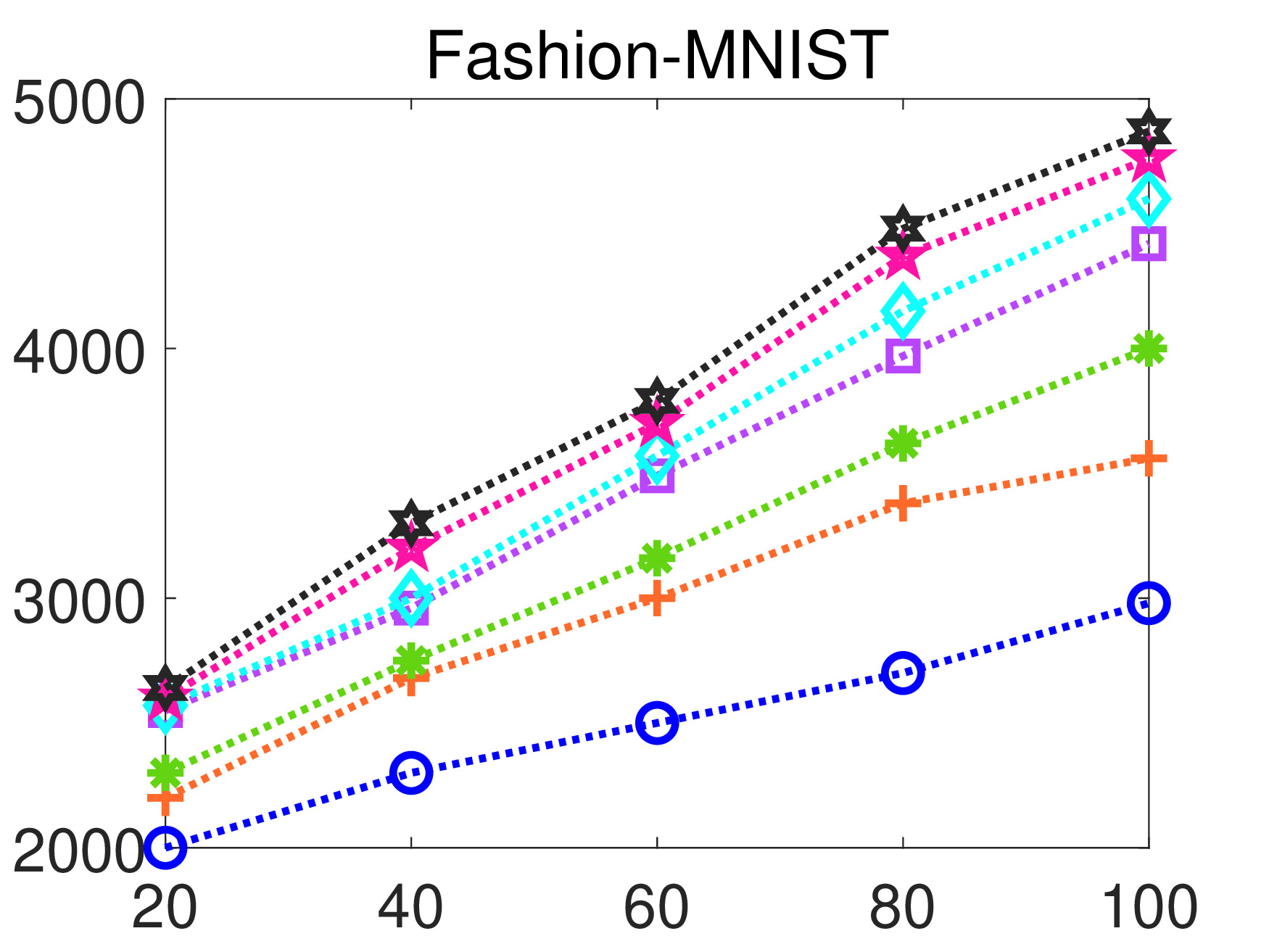}%
\label{fig_communication_Fashion-MNIST_case}}
\hfil
\subfloat[]{\includegraphics[width=1.2in]{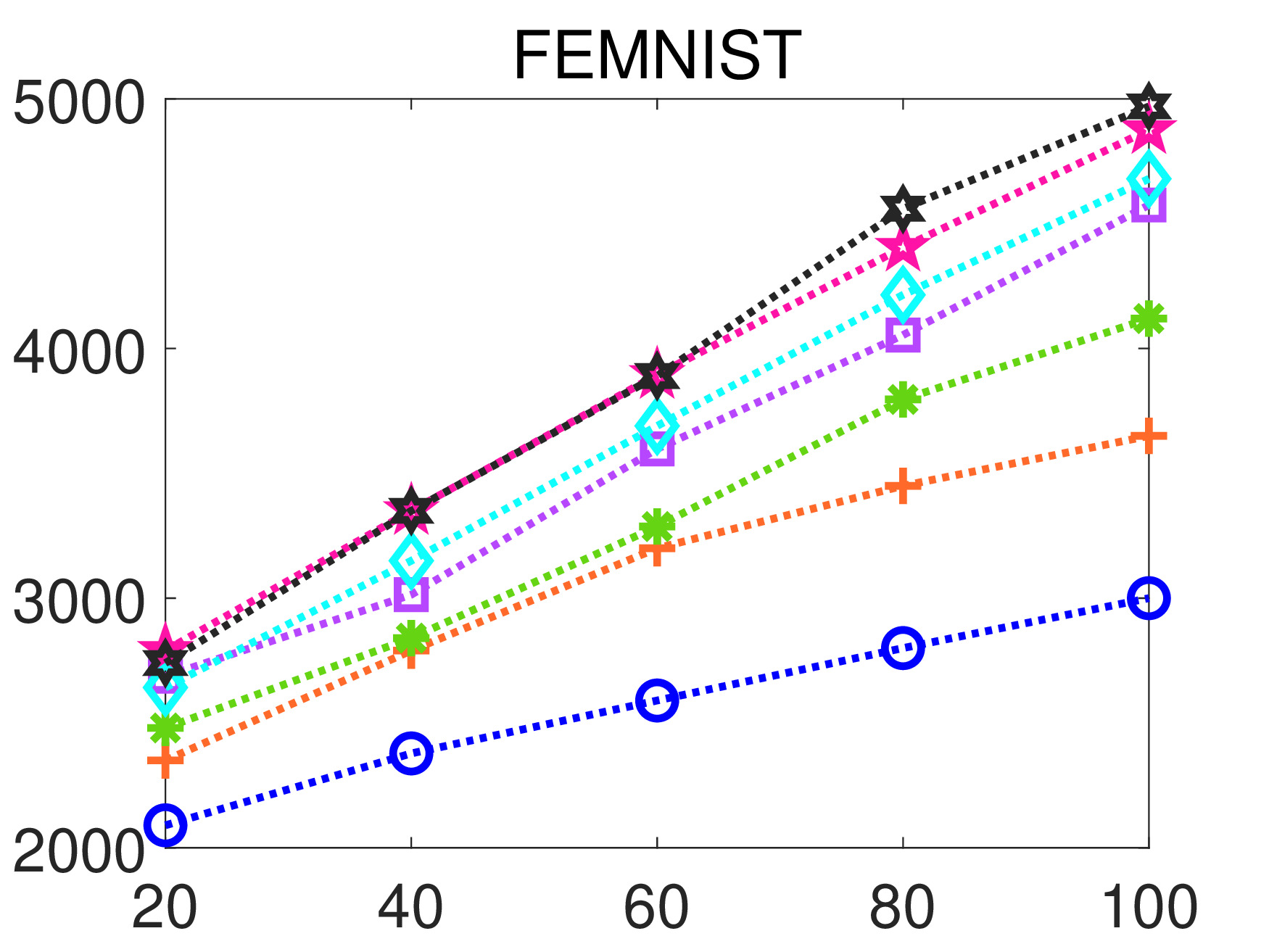}%
\label{fig_communication_FEMNIST_case}}
\hfil
\subfloat[]{\includegraphics[width=1.2in]{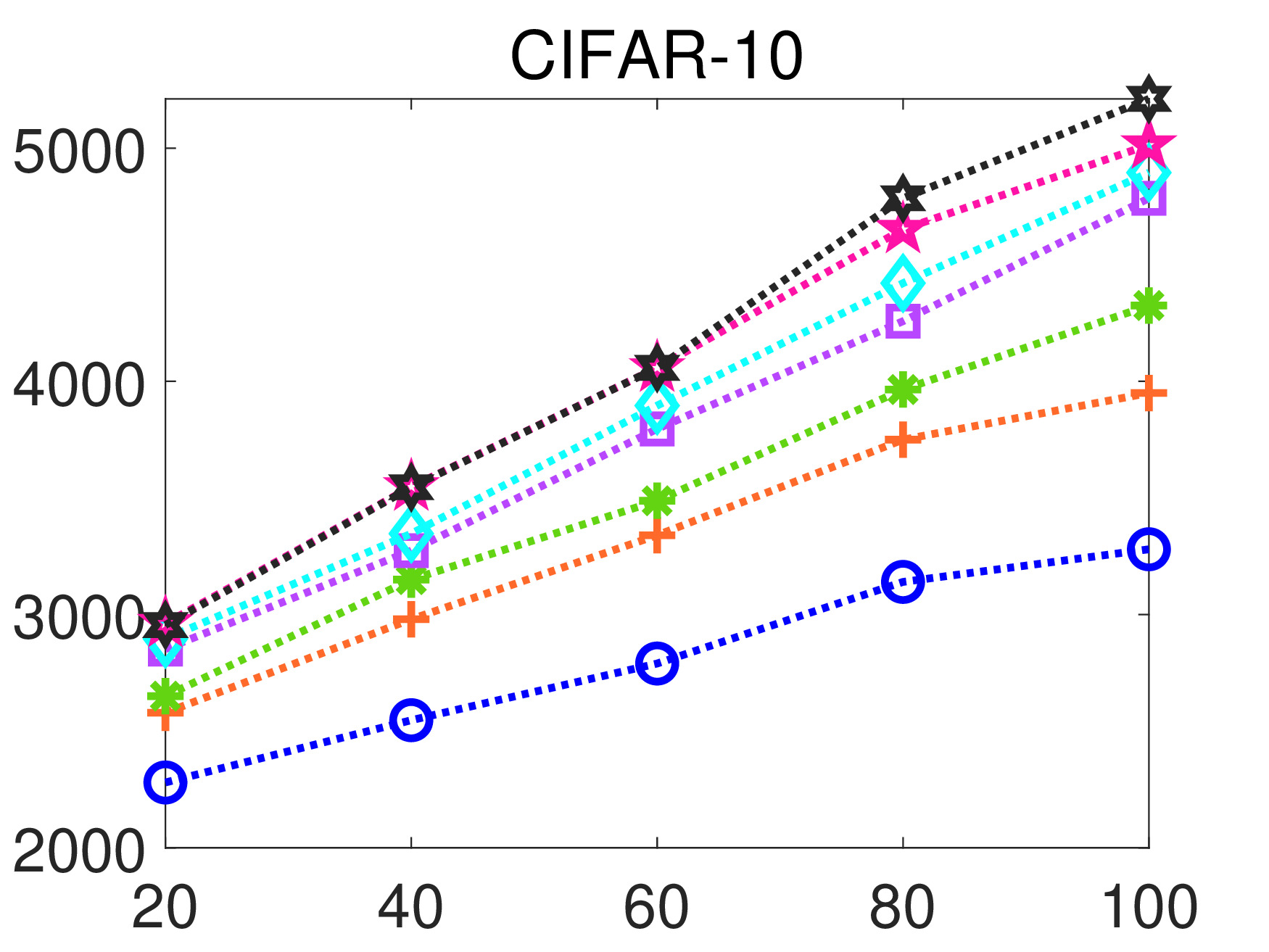}%
\label{fig_communication_CIFAr10_case}}
\hfil
\caption{Comparison of the communication cost between $\textsc{FedAnil+}$ and existing methods where the $\varepsilon$ is fixed at $30$ and $R=50$. ($\mu=20\%$).}
\label{fig_ALL_communication}
\end{figure*}

{\bf{\textsc{Computation Overhead.}}} As demonstrated in Fig. \ref{fig_computation_Sent140_case}, Fig. \ref{fig_computation_Fashion-MNIST_case}, Fig. \ref{fig_computation_FEMNIST_case}, and Fig. \ref{fig_computation_CIFAR10_case}, with the increase in number of enterprises, the $\textsc{FedAnil+}$ computation cost has increased. All approaches have been compared by increasing the number of enterprises from \emph{$20$} to \emph{$100$}. The computation cost also increases with the growth of enterprises. Because in the training phase, each local enterprise injects separate computations. This metric is intended to demonstrate the computation cost of the techniques employed. Based on the results presented in Fig. \ref{fig_ALL_computation}, the $\textsc{FedAnil+}$ model has low computational requirements due to avoiding heavy computational operations. In contrast, the {\textsc{STC}}, {\textsc{CFL}}, and {\textsc{RFA}} approaches utilized computationally heavy operations, such as sparse ternary encoding, aggregation oracles, and lossy encryption. On the other hand, the $\textsc{FedAnil+}$ has a higher computational cost than the {\textsc{FedAvg}} due to its use of quantization and homomorphic encryption operations. The low computational of the {\textsc{FedAvg}} is because it performs fewer computations on the client side, resulting in low overall computation overhead. While the $\textsc{FedAnil+}$ model has low computational overhead, it does not demonstrate better performance compared to the {\textsc{FedAvg}}.
\begin{figure*}[!t]
\centering
\subfloat[]{\includegraphics[width=1.2in]{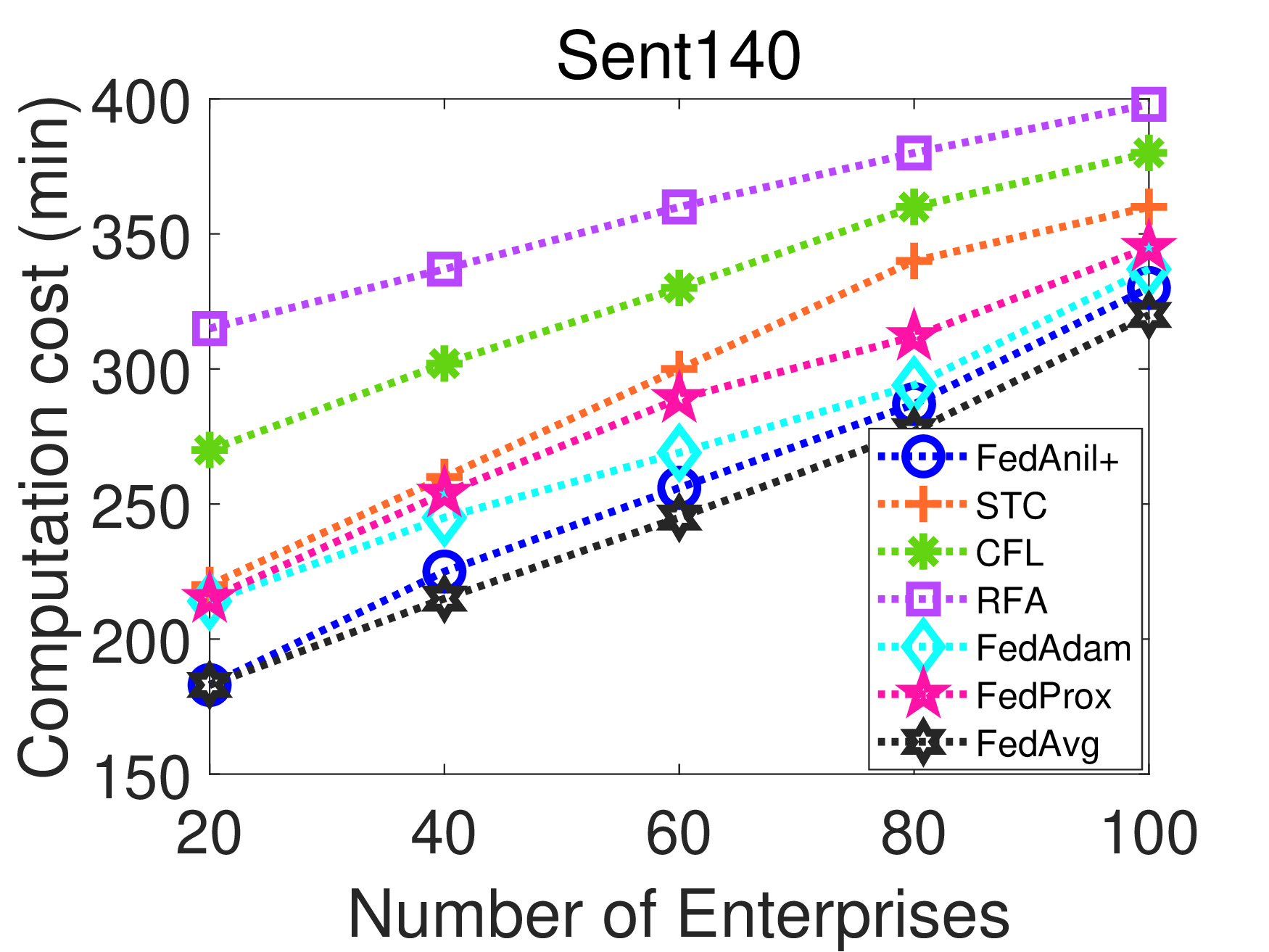}%
\label{fig_computation_Sent140_case}}
\hfil
\subfloat[]{\includegraphics[width=1.2in]{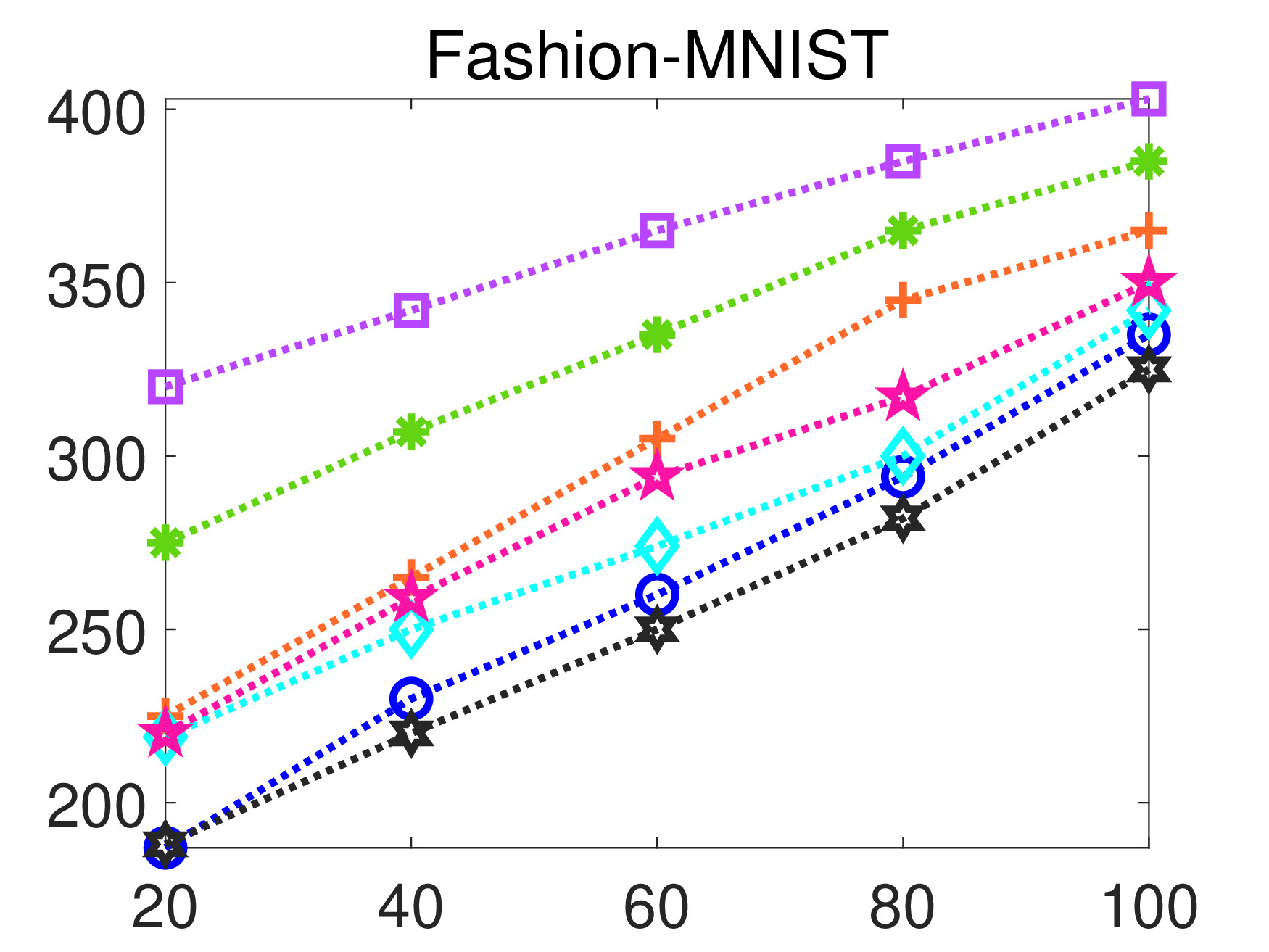}%
\label{fig_computation_Fashion-MNIST_case}}
\hfil
\subfloat[]{\includegraphics[width=1.2in]{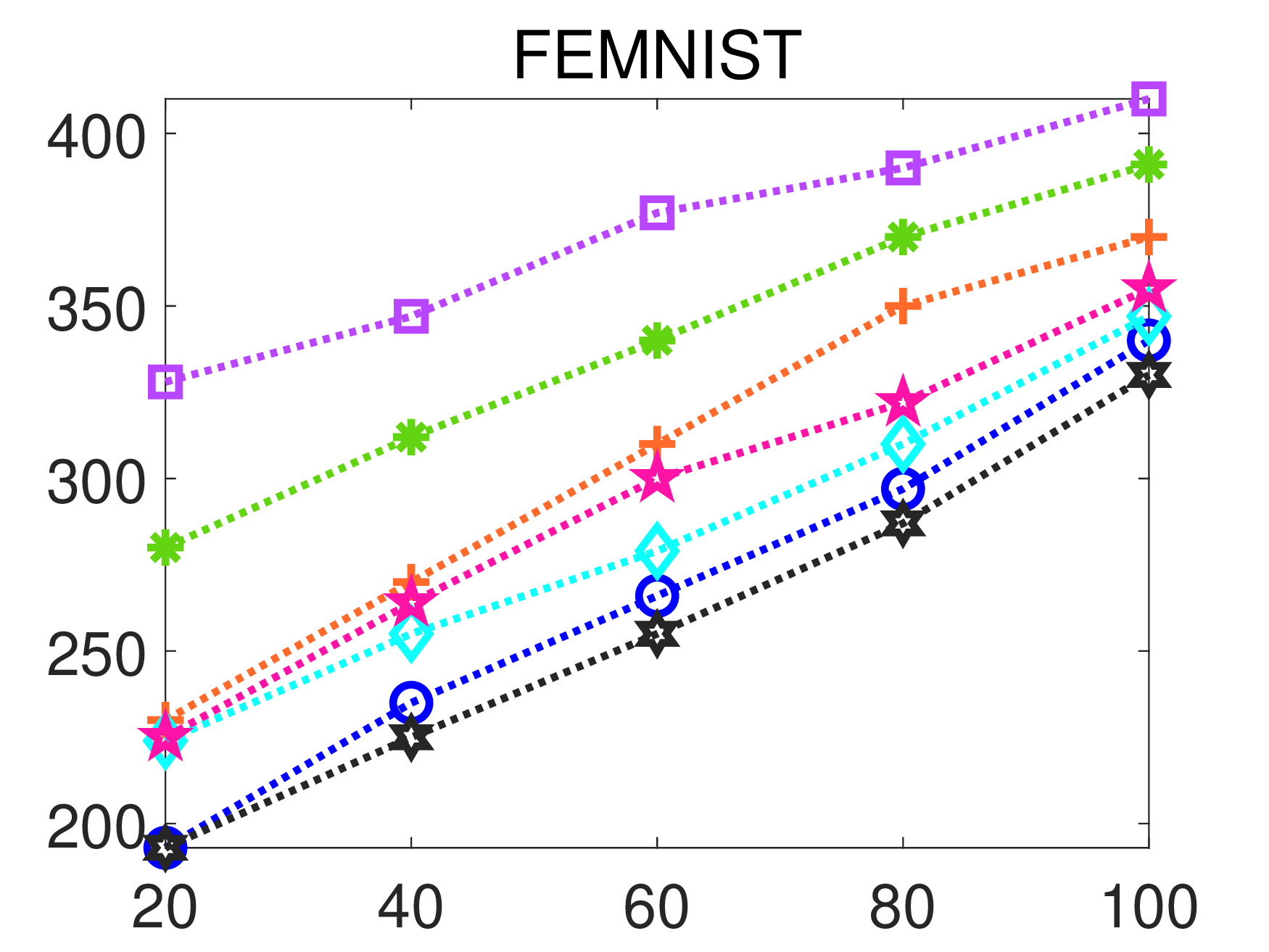}%
\label{fig_computation_FEMNIST_case}}
\hfil
\subfloat[]{\includegraphics[width=1.2in]{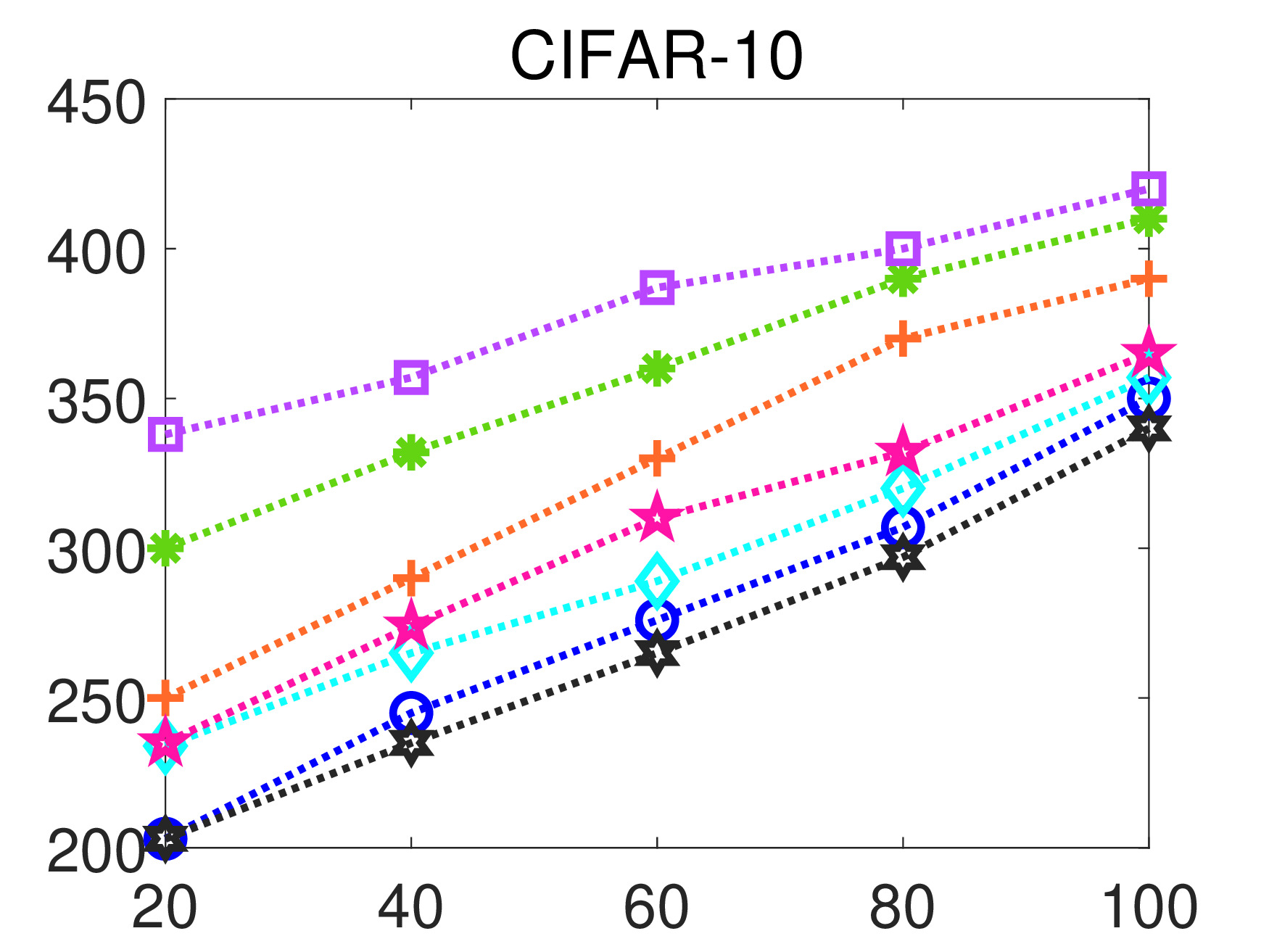}%
\label{fig_computation_CIFAR10_case}}
\hfil
\caption{Comparison of the computation cost between $\textsc{FedAnil+}$ and existing methods where the $\varepsilon$ is fixed at $30$ and $R=50$. ($\mu=20\%$).}
\label{fig_ALL_computation}
\end{figure*}

{\bf{\textsc{The $\textsc{FedAnil+}$ resistance to inference attacks.}}} To calculate the Gradient Matching Loss (GML), the L-BFGS model \cite{ref70} is adopted as an optimizer in $\textsc{FedAnil+}$. The difference between fake adversary-generated and real samples is represented by \emph{GML}. According to \cite{ref71}, any \emph{GML} value greater than $0.15$ will not leak any information. Conversely, the smaller the \emph{GML} value is than $0.15$, the more information is leaked. As Fig. \ref{fig_Comparission3Method3} shows, when the total number of rounds is set to $100$, in the $\textsc{FedAnil+}$ model and the execution of the 20th round, the value of the \emph{GML} is $0.17$. This means the difference between the real and fake parameters loss function is $0.17$. Continuing and repeating the number of rounds from $20$ to $100$, $\textsc{FedAnil+}$ has the same \emph{GML} value of $0.17$ due to its gradient leakage resistance approach and does not leak any information ({\bf{No Leak}}). Therefore, this non-leakage of parameters will be preserved until round $100$. In particular, the reason for the robustness of the $\textsc{FedAnil+}$ is a \emph{CKKS-FHE} technique, in which the encrypted local enterprise's models are aggregated without decryption on the server side. This makes the intruder not understand the content of the parameters. \emph{GML} value is $0.13$, $0.10$, and $0.09$ for STC, CFL, and RFA methods, respectively, which leaked some image's pixels, and due to the lack of a strong privacy protection approach, this leakage was repeated up to round $100$ ({\bf{Leak with artifacts}}). The \emph{GML} for FedAdam, FedProx, and FedAvg methods have values of $0.04$, $0.025$, and $0.015$, respectively. This means the methods in this \emph{GML} have many parameter matches, and the attacker's dummy data is much closer to the original data. Therefore, these methods have deep leakage and do not preserve the privacy of parameters ({\bf{Deep Leakage}}). By repeating the number of rounds from $20$ to $100$, the same deep leakage occurred, and a higher percentage of the real parameters information matches the fake parameters information by the attacker. The \emph{GML} value remains constant from round $20$ onwards because none of the approaches could prevent this recovery and leakage due to the lack of a robust privacy protection approach. Thus, the attacker was able to recover more real parameters.
\begin{figure}[!t]
	\centering
	\includegraphics[width=3.5in]{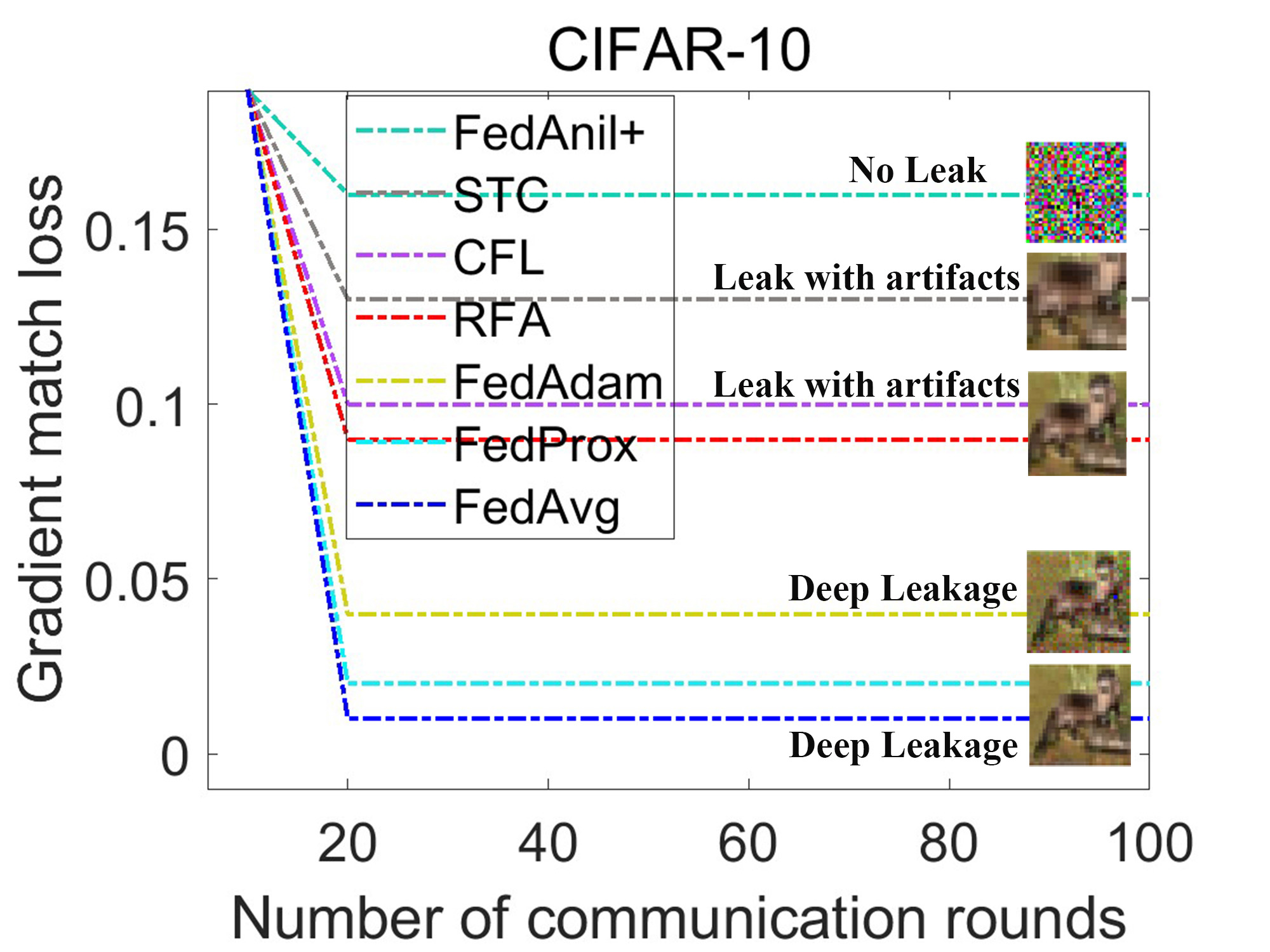}
\caption{\emph{GML} between $\textsc{FedAnil+}$ and existing methods where the $\varepsilon$ is fixed at $30$ and $R=100$. ($\mu=20\%$).}
	\label{fig_Comparission3Method3}
\end{figure}

\section{Conclusions and Future Work}\label{CC1}
This paper proposed a lightweight model named $\textsc{FedAnil+}$. In particular, the main innovation of $\textsc{FedAnil+}$, in addition to alleviating the privacy concern, is reducing the model's size and addressing the data type distribution skew. Our simulation results validate that the $\textsc {FedAnil+}$ improvements over existing approaches regarding accuracy, communication, and computation overhead. Moreover, the convergence analysis showed that the $\textsc{FedAnil+}$ model converges to the set of optimal model parameters. In future work, we will focus on three non-IID data distribution skews, i.e., Feature, Label, and Data type.

\vspace{-.3700in}

\begin{IEEEbiography}[{\includegraphics[width=1in,height=1.25in,clip,keepaspectratio]{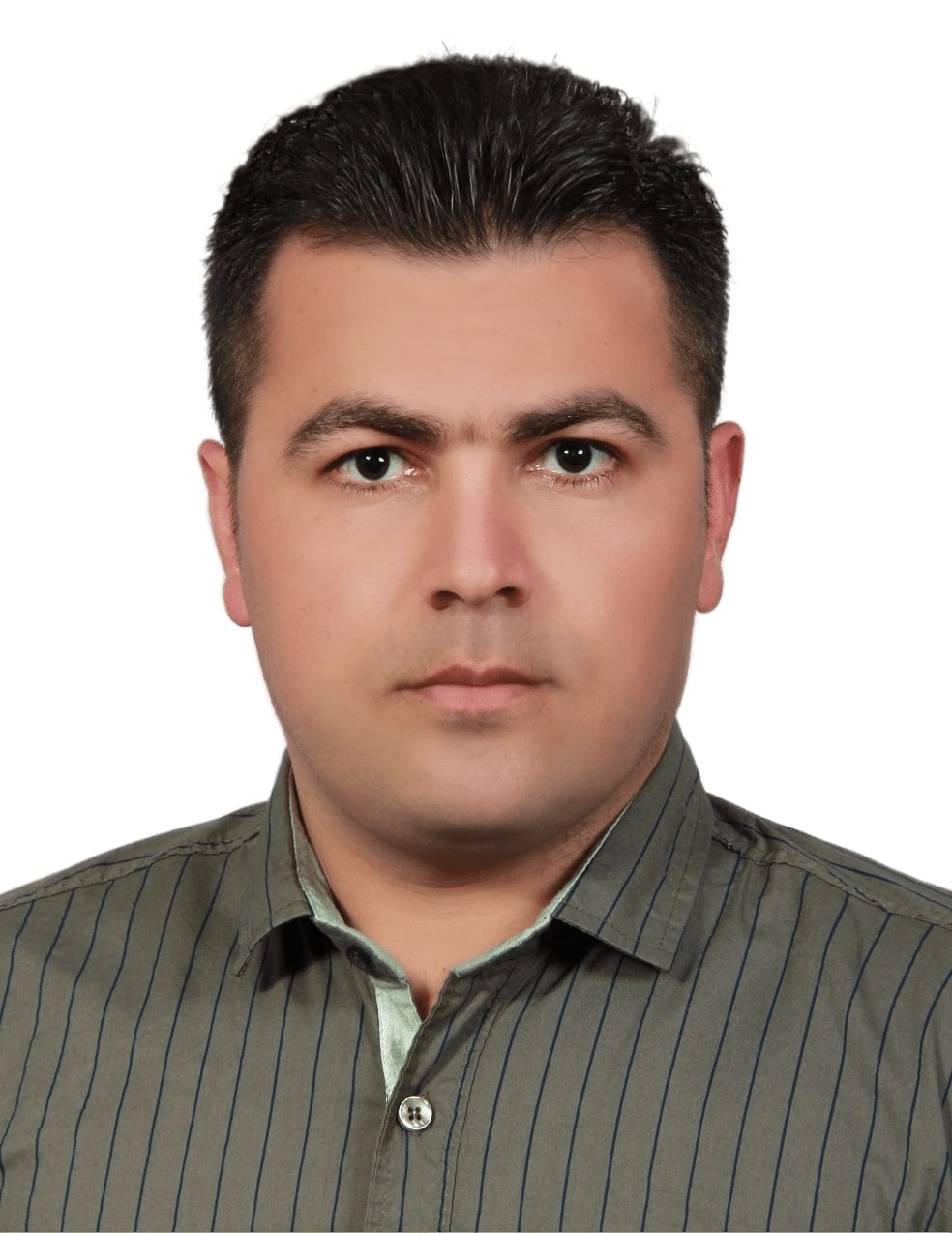}}]{Reza Fotohi} is a Ph.D. Candidate in the Faculty of Computer Science and Engineering, Shahid Beheshti University. His research interests include Privacy-Preserving Federated Learning (PPFL). Furthermore, he has published several papers in security domains in highly-ranked journals. His papers have over $1655$ citations with a $27$ h-index and $33$ i10-index. Also, he is recognized as being among the World's Top $2$\% of Scientists (Stanford University Ranking, $2021$ \& $2022$, \href {https://elsevier.digitalcommonsdata.com/datasets/btchxktzyw/3}{\color{blue}{Link}}). Reza Fotohi has served as a reviewer of several journals, such as IEEE Communications Surveys and Tutorial, IEEE Internet of Things Journal, IEEE Transactions on Aerospace and Electronic Systems, IEEE Transactions on Artificial Intelligence, IEEE Transactions on Cognitive Communications and Networking, IEEE Transactions on Reliability, ACM Transactions on Privacy and Security, Applied Soft Computing, Artificial Intelligence Review, Computers \& Security, etc. (\href {https://orcid.org/0000-0002-1848-0220}{\color{blue}{ORCID Link}}).
\end{IEEEbiography}

\vspace{-.3700in}

\begin{IEEEbiography}[{\includegraphics[width=1in,height=1.25in,clip,keepaspectratio]{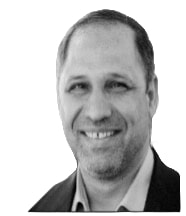}}]{Fereidoon Shams Aliee} received his Ph.D. in Software Engineering from the Department of Computer Science, Manchester University, in 1996 and his M.S. from the Sharif University of Technology, in 1990. His major interests are Software Architecture, Enterprise Architecture, Service-oriented Architecture, and Software Engineering. He is currently a Professor at the Shahid Beheshti University. Also, Dr. Shams is heading a research group, namely SOEA Lab, \href {https://soea.sbu.ac.ir/}{\color{blue}{Link}} at Shahid Beheshti University. (\href {https://orcid.org/0000-0002-9038-1577}{\color{blue}{ORCID Link}}).
\end{IEEEbiography}

\vspace{-.3700in}

\begin{IEEEbiography}[{\includegraphics[width=1in,height=1.25in,clip,keepaspectratio]{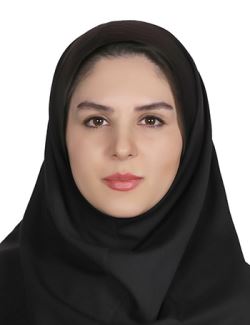}}]{Bahar Farahani} received her Ph.D. and Postdoctoral degrees in Computer Engineering from the University of Tehran, and Shahid Beheshti University, respectively. She is an assistant professor at Cyberspace Research Institute, Shahid Beheshti University. She authored several peer-reviewed Conference/Journal papers and book chapters on IoT, Big Data, and AI. Dr. Farahani has served as a Guest Editor of several journals, such as IEEE Internet of Things Journal (IEEE IoT-J), IEEE Transactions on Very Large Scale Integration Systems (IEEE TVLSI), and Elsevier Information Systems. (\href {https://orcid.org/0000-0002-7016-6853}{\color{blue}{ORCID Link}}).
\end{IEEEbiography}
\vfill
\end{document}